\documentclass[aps, pre, twocolumn, amsmath, superscriptaddress,showkeys,showpacs]{revtex4-1}
\usepackage[colorlinks=true,
linkcolor=red,
urlcolor=black,
citecolor=blue]{hyperref}
\usepackage{graphicx}
\usepackage{epstopdf}
\usepackage{dcolumn}
\usepackage{bm}
\usepackage{soul}

\usepackage{color}
\definecolor{darkgreen}{rgb}{0.1,.6,.1}
\definecolor{greenblue}{rgb}{0.0,.1,.4}
\usepackage[normalem]{ulem}

\begin{document}


\title {Instabilities in quasiperiodic motion lead to intermittent large-intensity events in Zeeman laser}
\author{S. Leo Kingston}
\thanks{kingston.cnld@gmail.com}	
\affiliation{Division of Dynamics, Lodz University of Technology, 90-924 Lodz, Poland}
\author{Arindam Mishra}
\affiliation{Division of Dynamics, Lodz University of Technology, 90-924 Lodz, Poland}
\author{Marek Balcerzak}
\affiliation{Division of Dynamics, Lodz University of Technology, 90-924 Lodz, Poland}
\author{Tomasz Kapitaniak}
\affiliation{Division of Dynamics, Lodz University of Technology, 90-924 Lodz, Poland}
\author{Syamal K. Dana}
\affiliation{Division of Dynamics, Lodz University of Technology, 90-924 Lodz, Poland}
\affiliation{National Institute of Technology, Durgapur 713209, India}



\date{\today}

\begin{abstract}
We report intermittent large-intensity pulses that originate in Zeeman laser  due to instabilities in quasiperiodic motion, one route follows  torus-doubling to chaos and another goes via quasiperiodic intermittency in response to variation in system parameters. The quasiperiodic breakdown route to chaos via torus-doubling is well known, however, the laser model shows intermittent large-intensity pulses for parameter variation beyond the chaotic regime.  During  quasiperiodic intermittency, the temporal evolution of the laser shows intermittent chaotic bursting episodes intermediate to the quasiperiodic motion instead of periodic motion as usually seen during the Pomeau-Manneville intermittency. The intermittent bursting appears as occasional large-intensity events. In particular, this quasiperiodic intermittency has not been given much attention so far from the dynamical system perspective, in general. In both the cases, the infrequent and recurrent large events show non-Gaussian  probability distribution of event height extended beyond a significant threshold with a decaying probability confirming rare occurrence of large-intensity pulses. 

\end{abstract}

\maketitle

\section{Introduction}

\par Instabilities in lasers have long been investigated to explain
the formation of chaos \cite{Arecchi, Arecchi1, Arecchi2, Roy, Junji}, which originates, in many cases via Pomeau-manneville (PM) intermittency \cite{PM}. However, chaotic pulses
were accompanied by occasional large-intensity pulses in a range of parameters, which was not recognized as a  distinctly different phenomenon.  Much later, large-intensity traveling pulses in a two-dimensional array of frequency-disordered laser oscillators were reported \cite{Roy1} as local coherent large-amplitude excitations. Around the same time, optical rogue waves as rare giant pulses were reported \cite{Solli, Bonatto, Pisarchik} that originated due to deterministic or stochastic nonlinear processes. In recent time,  intermittent large-intensity pulses have been reported in many laser systems, optically injected cavity \cite{Montina}, solid-state laser \cite{Granese},  semiconductor laser \cite{Rimoldi}, and $\mathrm{CO_2}$ laser \cite{Cristian-CO2} that are recognized as extreme events and different in character from nominal chaos with limited amplitude  below a threshold height. The threshold  height is determined by a statistical measure \cite {Dysthe, Arnob_Phys_Report, Mercier} from time evolution of an observable for a long time. Optically injected semiconductor lasers manifest rare ultra-intensity pulses in a narrow range of parameter region.
The appearance and termination of extremely large-intensity pulses or events were implemented by feedback control in semiconductor or diode lasers \cite{Reinoso, Mercier}.  In deterministic laser systems \cite{Zamora-Munt, Metayer}, interior-crisis-induced intermittency \cite{Grebogi, Rakshit, kingston} is mostly found as a nonlinear process responsible for occasional large-intensity pulses. Noise-induced attractor-hopping in multistable systems \cite{Pisarchik} also triggers rare large-intensity events. In spatially extended microcavity laser,  evidence of extremely large events was found  \cite{Coulibaly, Clerc, Clerc1} that originated due to spatio-temporal chaos and intermittency. 
\par An obvious question arises as to whether these processes are exhaustive or there exist other possible sources of instabilities that may induce intermittent large events. The answer lies in the search of extraordinary large events that may emerge in dynamical systems, in general. In recent years, various dynamical models had been investigated \cite{Lucerini, deOliveira, kingston, Suresh_2018, Karnatak_2014, Mishra_2018, Ray_2019, Chang, Mohamad, Ott, Sayantan, Sayantan1, Majhi, Arnob} and real-time laboratory experiments \cite{Zamora-Munt, Reinoso,  Clerc1, kingston, Mishra_2018} were done where similar occasional large-amplitude events were recorded. Attempts have been made to discern the underlying mechanisms of the origin of such extremes and their statistical properties. A general perception has been developed \cite{kingston, Farazmand, kingston2020, Arnob_Phys_Report} that an  instability region may exist in the state space of a nonlinear system. The trajectory of the system may occasionally visit a close vicinity of the instability region and it is diverted to far away locations for a short duration, but returns to the nominal state after a short duration. The trajectory of the system otherwise remains confined, most of the time, in  the nominal state within a bounded volume of the state space. The occasional large excursions form intermittent large-amplitude events; the source of instabilities only differs from system to system. The large-amplitude events usually follow a non-Gaussian distribution with a tail  (long-, heavy-tail) and, in some special cases, a dragon-king-like distribution \cite{deOliveira, Mishra_2018, Mishra2020} when the  large events are outliers to a power law. A complete understanding of  the dynamical processes involved in the origin of the rare  large-amplitude events in any system is an essential  task for developing  an appropriate technique for early forecasting \cite{Mohamad}. 
\par Besides system-specific several sources of instabilities  \cite{Suresh_2018, Arnob_Phys_Report, Ott, Sayantan, Karnatak_2014, Arnob, Lehnertz} as reported, in the literature, we identify three fundamental sources  in dynamical systems, single systems, coupled systems, and network of systems that create instability in the systems and may trigger intermittent large events, {\it interior crisis-induced intermittency} \cite{Grebogi, Reinoso, Rakshit}, {\it PM intermittency} \cite{PM, Mishra_2018, kingston}, and {\it breakdown of quasiperiodic (QP) motion} \cite{Mishra2020, Nicolis}. Interior-crisis-induced intermittent large events occur due to a collision of a chaotic trajectory with a stable manifold of a saddle point or a saddle orbit that coexists in the state space of a system. This particular phenomenon has been observed in many model systems as well as experiments in lasers as mentioned above,  which appears after the origin of chaos via a period-doubling cascade in response to  a parameter variation. As mentioned, in the beginning, PM intermittency may lead to occasional large events, which manifested as occasional chaotic bursting (turbulent phase) intermediate to almost periodic oscillation (laminar phase) in the time evolution of a state variable of a system. In some systems, chaotic bursts occasionally appear with extremely large amplitude compared to the amplitude of the periodic flow in the laminar phase. This particular situation was demonstrated in a forced Li\'enard system \cite{kingston} and a coupled neuron model under chemical synaptic interaction \cite{Mishra_2018, Mishra2020}. The breakdown of QP motion via torus-doubling is another route to chaos that may lead to occasional large events \cite{Mishra_2018}. The possibility of extreme behavior via breakdown of QP motion was predicted earlier \cite{Nicolis} in a map during a study of extreme events using statistical approaches; however, there the dynamics of large events was not looked into. 
\par We revisit the dynamics of the Zeeman laser model with a large cavity anisotropy as reported earlier \cite{Puccioni, Abraham, Redondo, Redondo1} using numerical simulations. This study reported \cite{Redondo, Redondo1} origin of chaos via breakdown of quasiperiodicity. However, they did not pay attention to the complexity of dynamics beyond chaos. We scrutinize the parameter space in the laser model for the origin of chaos and beyond and,  locate two distinctly different instability sources.  
We discern the sources of instability in parameter space: (1) breakdown of QP motion to chaos via torus-doubling followed by another state   with a tuning of a system parameter when  intermittent large-intensity pulses originate, (2)  QP  intermittency, which is a relatively unknown phenomenon so far. The time evolution of QP intermittency shows a {\it laminar phase} of QP motion instead of a periodic motion as usually seen during PM intermittency \cite{PM} while the {\it turbulent phase} consists of chaotic bursting as usual. We explain the two nonlinear deterministic processes, so far remain unrecognized, to demonstrate
the origin of intermittent large-intensity pulses in the laser model.
\par We organize the text as follows: 
The Zeeman laser model is presented in Sec.~II with phase diagrams in two-parameter plane to locate the  sources of instabilities that lead to occasional large events. The torus-breakdown of QP motion and the QP intermittency routes to intermittent large events are elaborated in Secs.~III and  Sec.~IV, respectively, with one-parameter bifurcation diagrams, a series of temporal evolution of system dynamics for a varying  parameter, return maps of local maxima, and their probability distributions. Finally, results are summarized with a conclusion in Sec. V 
\section{Zeeman Laser Model}
A monochromatic electric field interacts inside a ring cavity with a homogeneously
broadened medium that consists of two-level atoms with lower ($J = 0$) and upper ($J = 1$) levels \cite{Redondo, Redondo1}. With mean-field and rotating wave approximations, and assuming a perfect resonance between the cavity and atomic frequencies, the Maxwell-Bloch equations describe the two-level
Zeeman laser model in dimensionless form,
\begin{eqnarray}
\dot{E}_x &=&\sigma(P_x - E_x), \\ \nonumber
\dot{E}_y &=&\sigma(P_y - \alpha E_y), \\ \nonumber
\dot{P}_x &=&-P_x +E_x D_x+E_y Q, \\  \nonumber
\dot{P}_y &=&-P_y +E_y D_y+E_x Q, \\  \nonumber
\dot{D}_x &=&(r-D_x)-2(2E_x P_x+E_y P_y), \\  \nonumber
\dot{D}_y &=&(r-D_y)-2(2E_y P_y+E_x P_x), \\  \nonumber
\dot{Q} &=&-Q-(E_x P_y+E_y P_x).   \nonumber
\end{eqnarray}
The state variables $E_x$ and $E_y$ represent the linear polarization components of the electric field,  ($P_x, P_y$) and ($D_x, D_y$) are proportional to the polarization and atomic inversion, respectively, which is related to a transition $| J=1, J_i = 0\rangle \leftrightarrow |J=0\rangle$, and $Q$ is proportional to the coherence between the upper sub levels $| J=1, J_x = 0\rangle$ and $| J=1, J_y = 0\rangle$. The parameter $r$ denotes the incoherent pumping rate, $\sigma$ and $\alpha \sigma$ represent the cavity losses along the $x$ and $y$ directions, where $\alpha$ is the cavity anisotropy parameter.
\par
\begin{figure}
	\includegraphics[width=0.9\columnwidth]{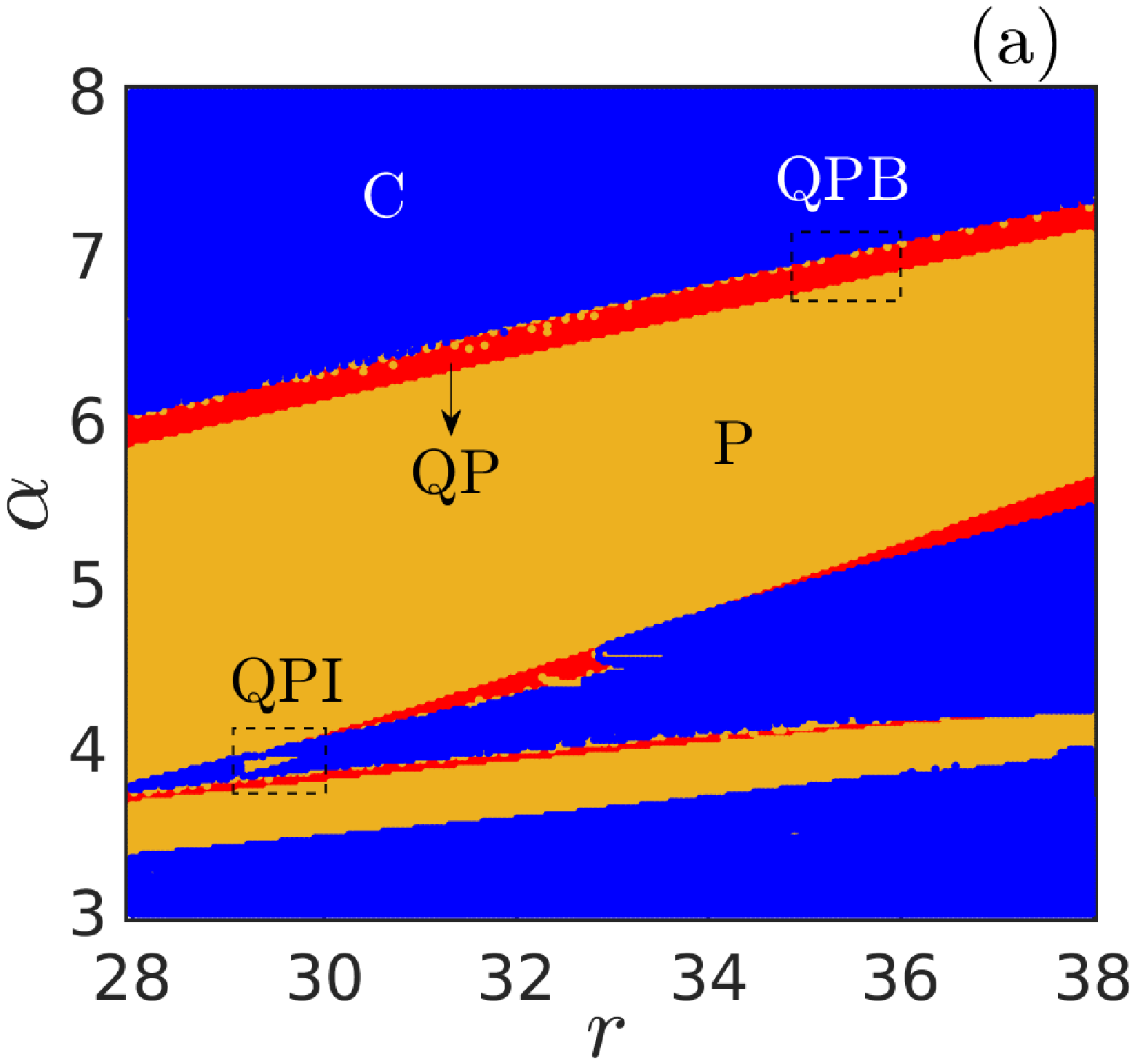} \\	\includegraphics[width=0.49\columnwidth]{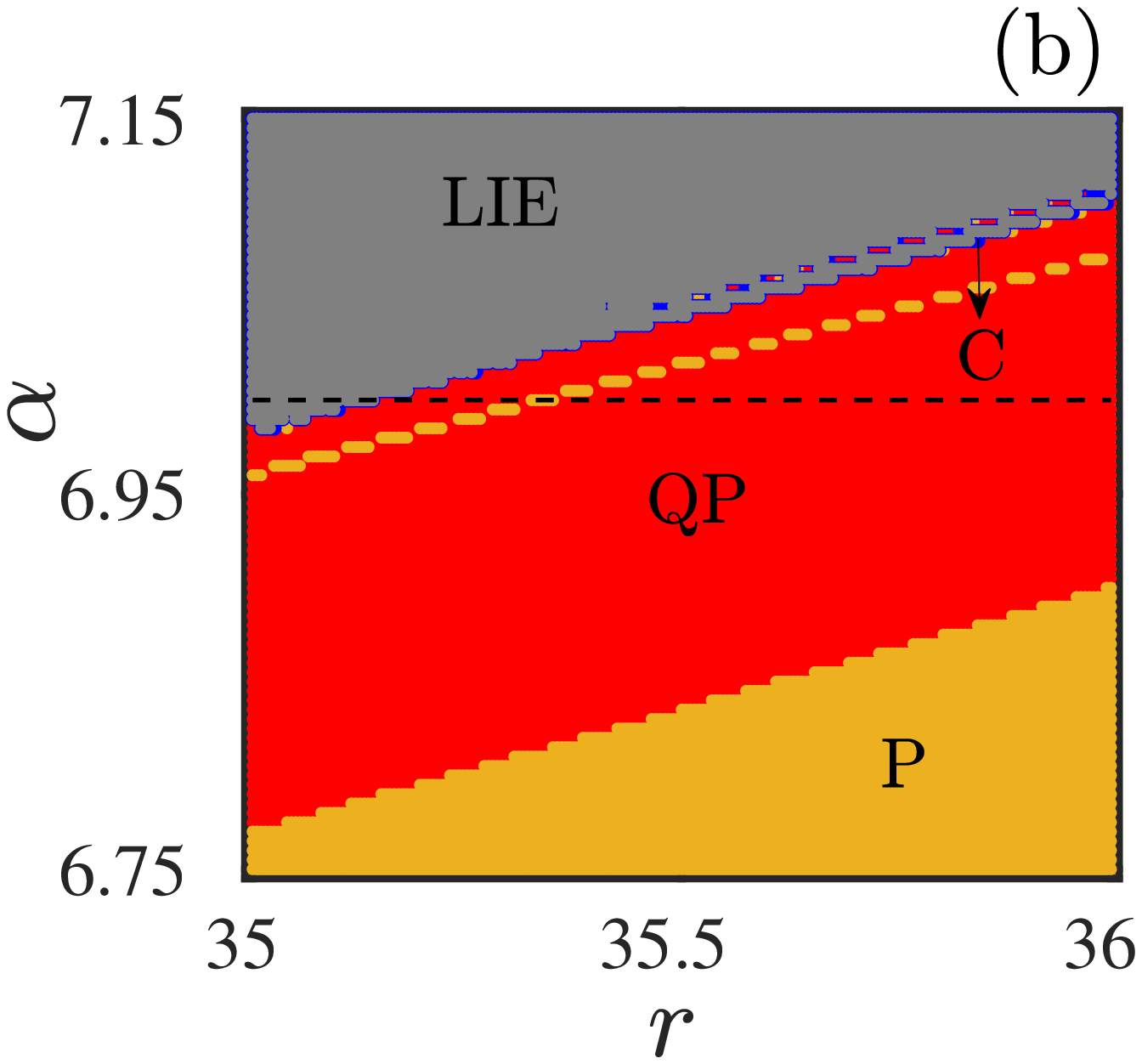}~	\includegraphics[width=0.49\columnwidth]{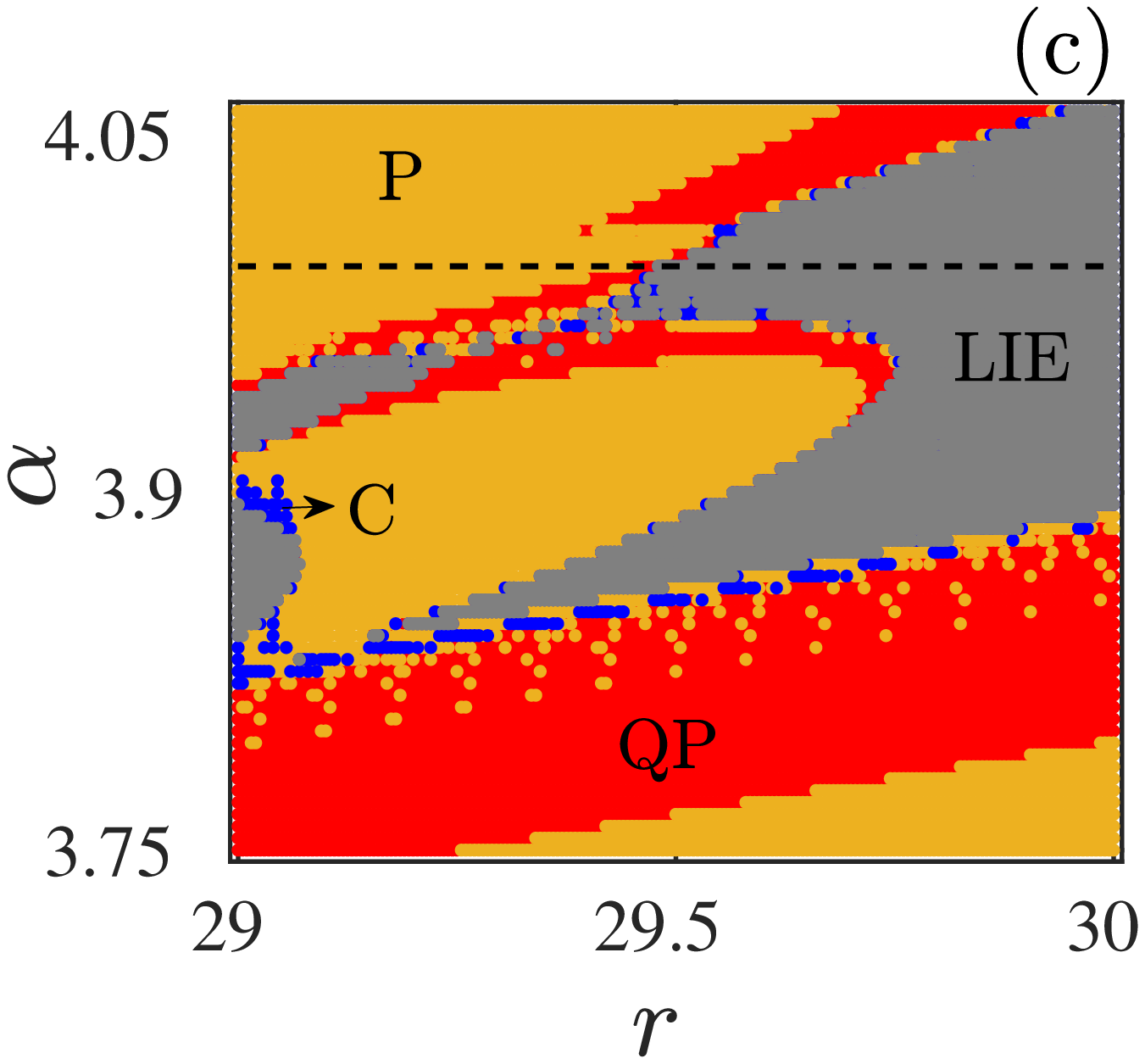} 
	\caption{Phase diagram in two-parameter plane of Zeeman laser model. (a) Blue region represents chaotic dynamics (C), quasiperiodic (QP) motion in red color, and yellow region represents periodic (P) state of the system. Zoomed versions of the (b) quasiperiodic breakdown (QPB) and (c) quasiperiodic intermittency (QPI) region [marked by dashed rectangles in (a)]. LIE region (gray color) is delineated when  any event is larger than a threshold height. Here $\sigma$ is fixed as 6.0.}
	\label{Zee:2ph}
\end{figure}

  In order for the variety of dynamics of the system to locate, in parameter space, as presented in Ref.~\cite {Redondo1}, we first  plot the phase diagram in the ($r - \alpha$) parameter plane for a fixed value of $\sigma$ = 6.0,  where the dynamics of each coordinate point is recognized by its respective Lyapunov exponents. The system manifests periodic (P in yellow), quasiperiodic (QP in red), and chaotic dynamics (C in blue) as shown in Fig.~\ref{Zee:2ph}(a). By a closer inspection of the chaotic region (blue), we locate two significantly disparate dynamical regions, in parameter space, quasiperiodic breakdown (QPB) and quasiperiodic intermittency (QPI) (marked by dashed rectangles), where the system exhibits intermittent large-intensity events (LIE) although they are identified earlier \cite{Redondo} as simply chaotic in nature. A positive value of the largest Lyapunov exponent can distinguish chaos, but failed to recognize the LIE, which is also chaotic in character. We discern the LIE from nominal chaos by the size of events and comparing them against a threshold height. If local maxima of laser intensity remain bounded for a long time below the threshold, then we call it nominal chaos. However, if some of the large-intensity peaks have height larger than the threshold, then we distinguish them as LIE. To delineate the LIE states, in parameter plane, we plot two additional phase diagrams (lower panels of Fig.~\ref{Zee:2ph}) that focus on the narrow range of the parameter plane close to the QPB and QPI regimes in Figs.~\ref{Zee:2ph}(b) and Fig.~\ref{Zee:2ph}(c), respectively. The QPB region shows a periodic regime (P, yellow), a quasiperiodic regime (QP, red), and the LIE region (gray), but with a very narrow chaotic region (C, blue) in between. QPI region also shows islands of periodic regime (yellow, P), quasiperiodic regime (QP, red), and chaotic regime (C, blue) with a sea of LIE state (gray). The narrow chaotic regimes (blue) in Figs.~\ref{Zee:2ph}(b) and \ref{Zee:2ph}(c), show no large events. 
Emergence of LIE is presented sequentially with a variation of $r$ along with their statistical properties, for the two distinct sources of instabilities, in the next sections.
\begin{figure}
\includegraphics[width=0.9\columnwidth]{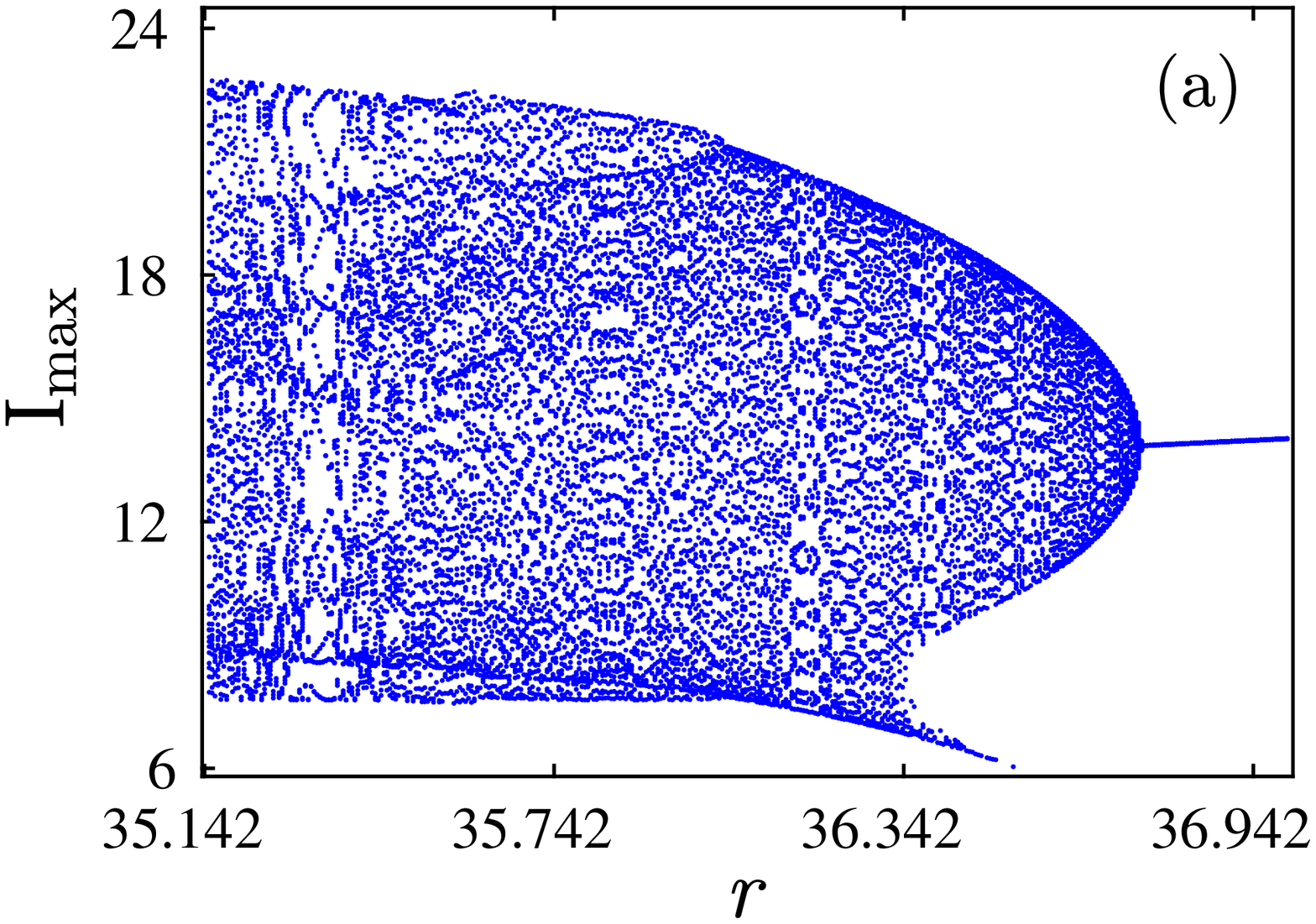} \\
 \includegraphics[width=0.95\columnwidth]{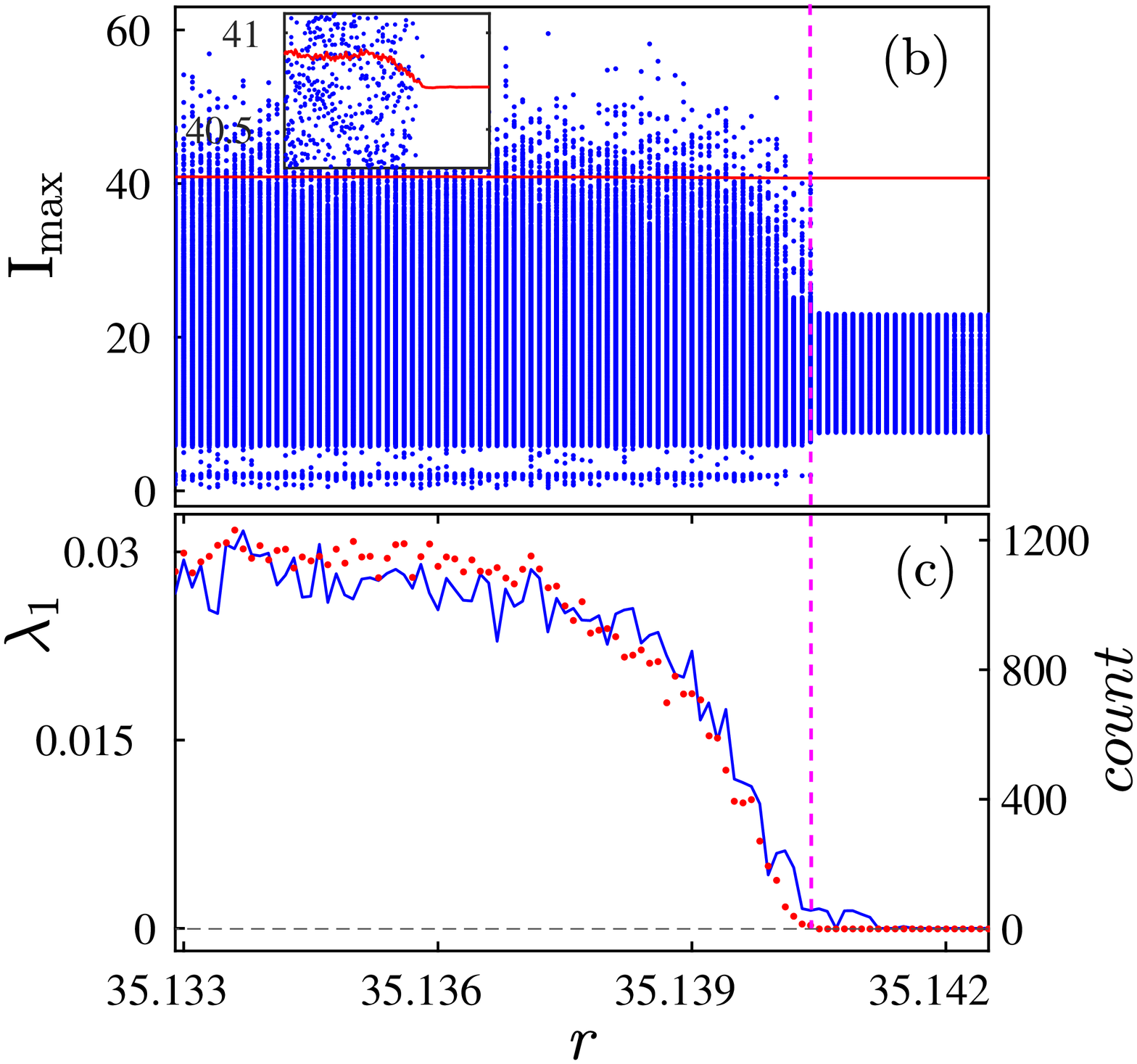}
	\caption{Bifurcation diagram of laser intensity against the pumping rate in Zeeman laser. Local maxima $I_{max}$ shows (a) a transition from periodic to QP motion and then (b) a sudden change in $I_{max}$ indicating emergence of LIE. The  transition to chaos via breakdown of QP is indicated by a transition of the largest Lyapunov exponent $\lambda_1$ from zero to a positive value of $\lambda_1$ at $r \approx 35.1405$ (c). LIE start appearing at a lower $r \le 35.1404$. The vertical dashed lines in panels (b) and (c) indicate the transition point at $r \approx 35.1404$ from nominal chaos to LIE. A horizontal line (red line) depicts the  $\mathrm{H_s}$= $\langle I_n\rangle$ + $6\sigma_{I}$. The number of extreme events varies with $r$ as plotted in red dots, showing the count at right-side scale (c).}
	\label{QPB:bif_lya}
\end{figure}
\begin{figure}
	\includegraphics[width=0.55\columnwidth]{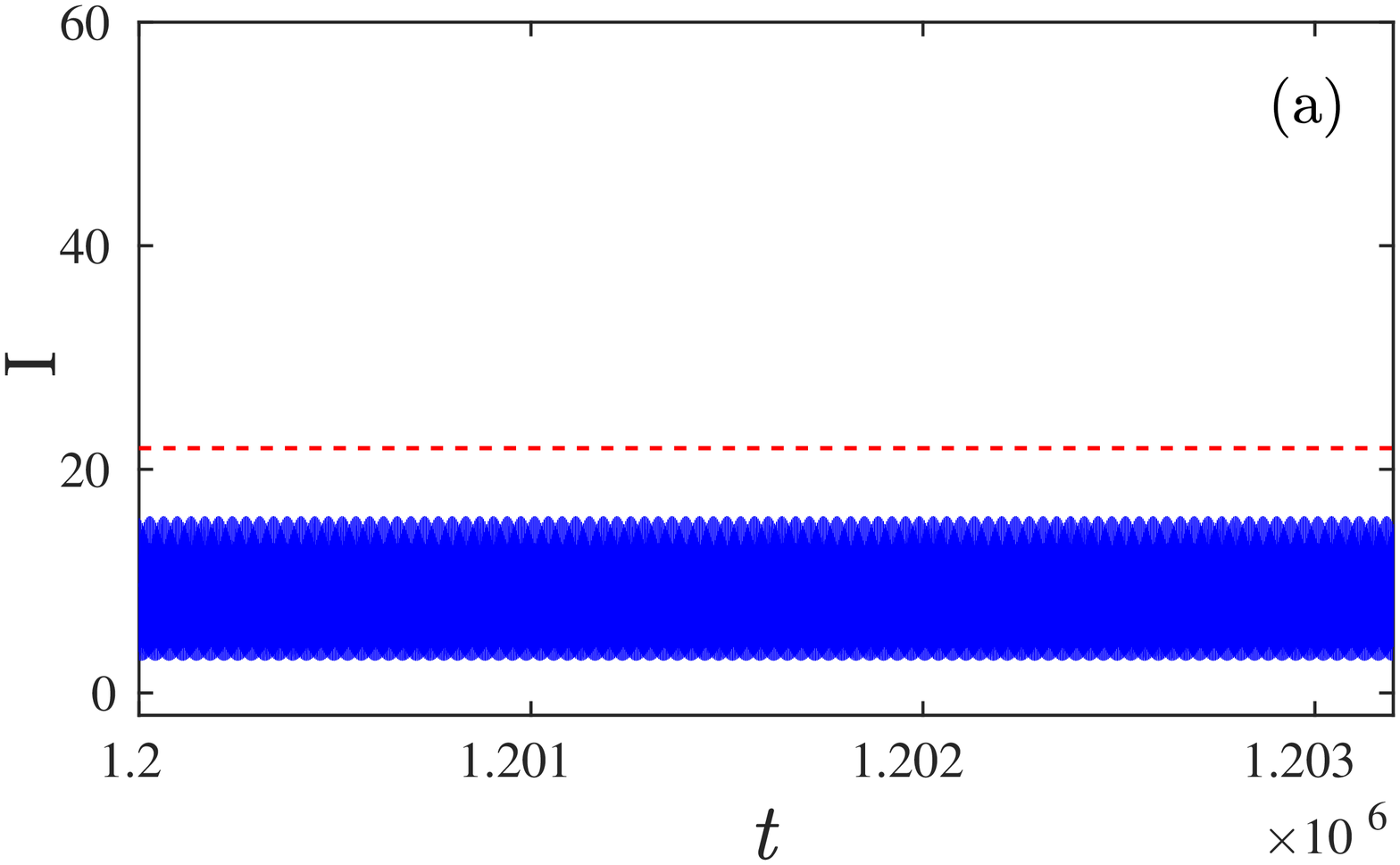}~\includegraphics[width=0.4\columnwidth]{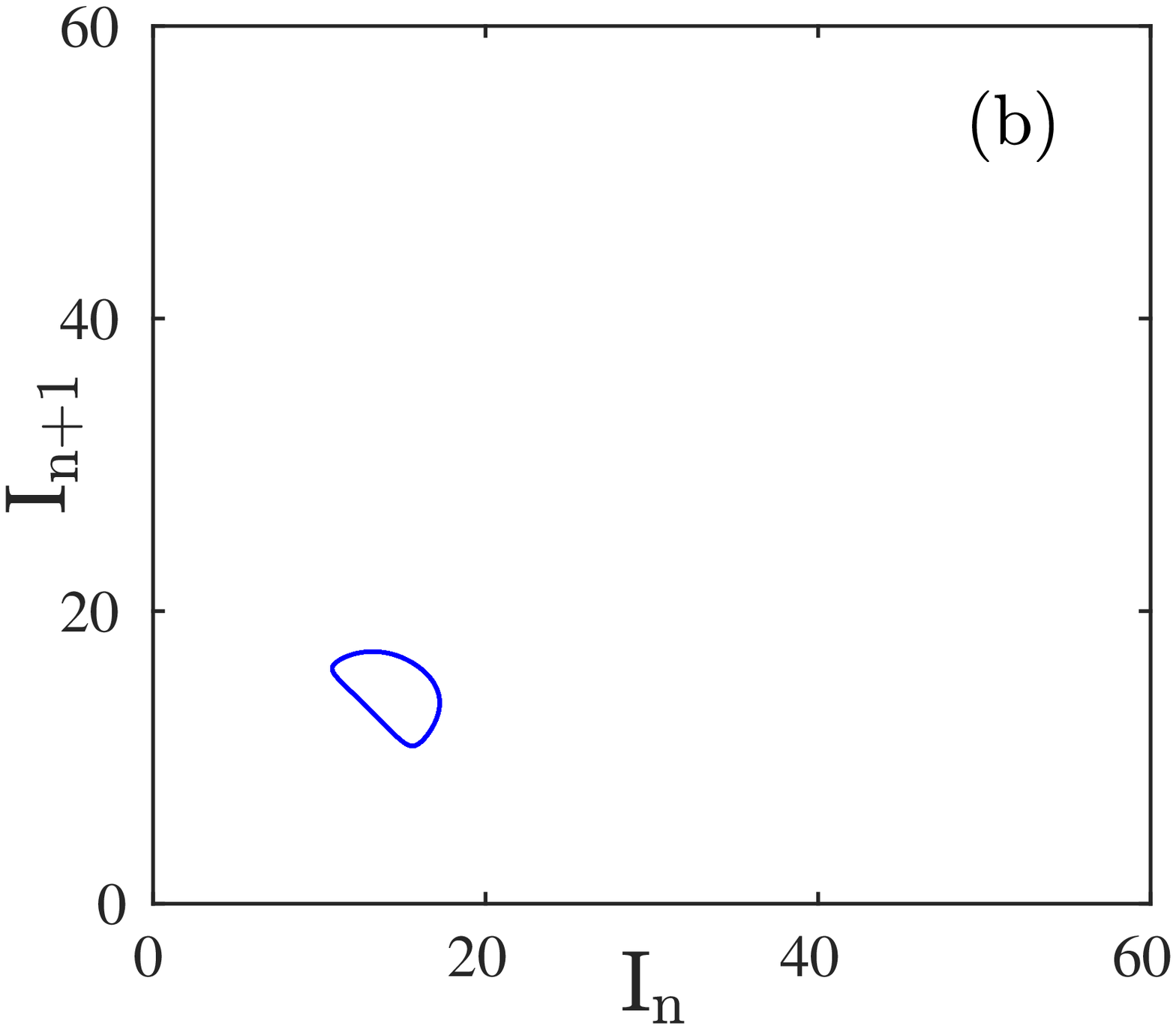} \\
	\includegraphics[width=0.55\columnwidth]{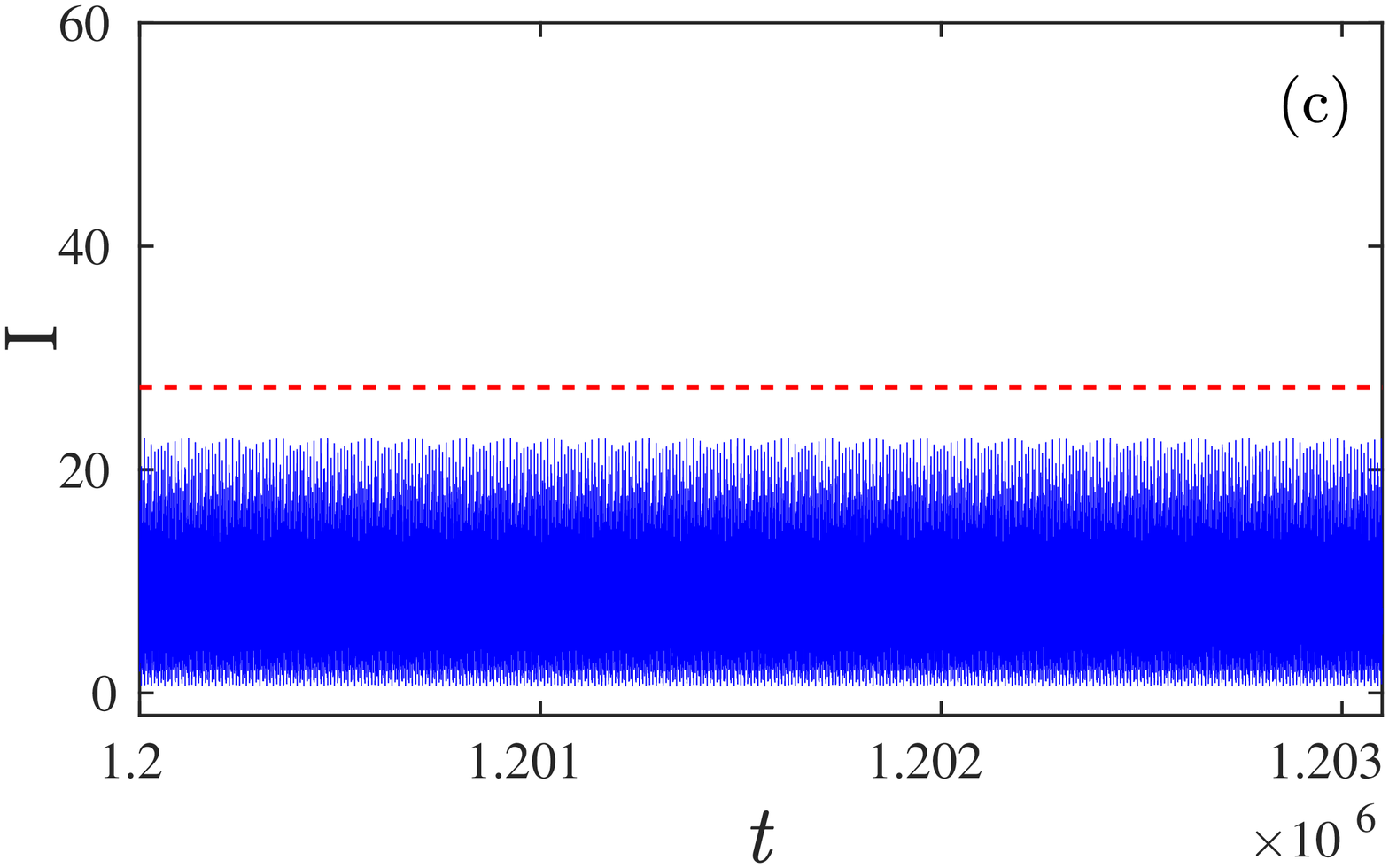}~\includegraphics[width=0.4\columnwidth]{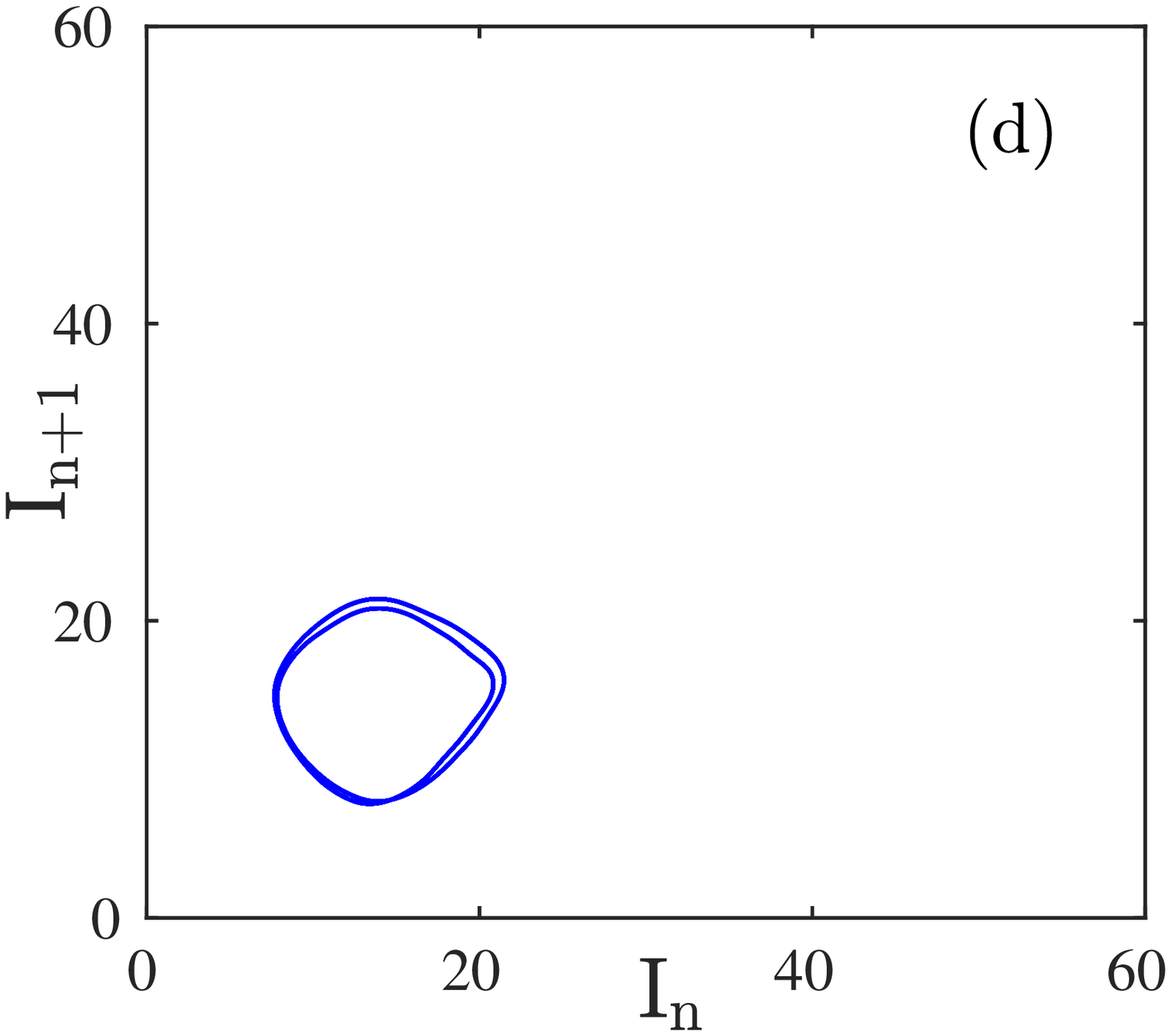}\\
	\includegraphics[width=0.55\columnwidth]{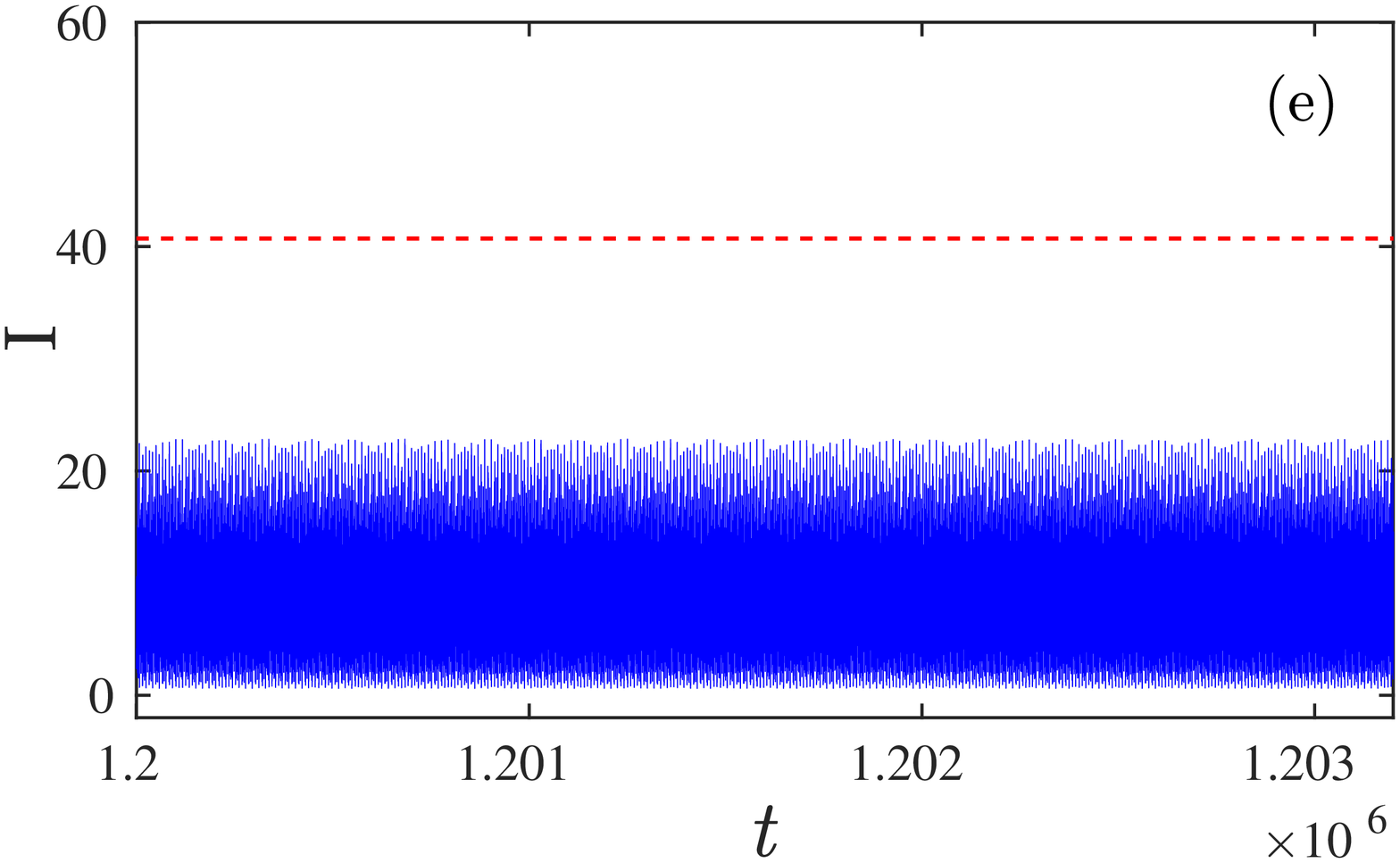}~\includegraphics[width=0.4\columnwidth]{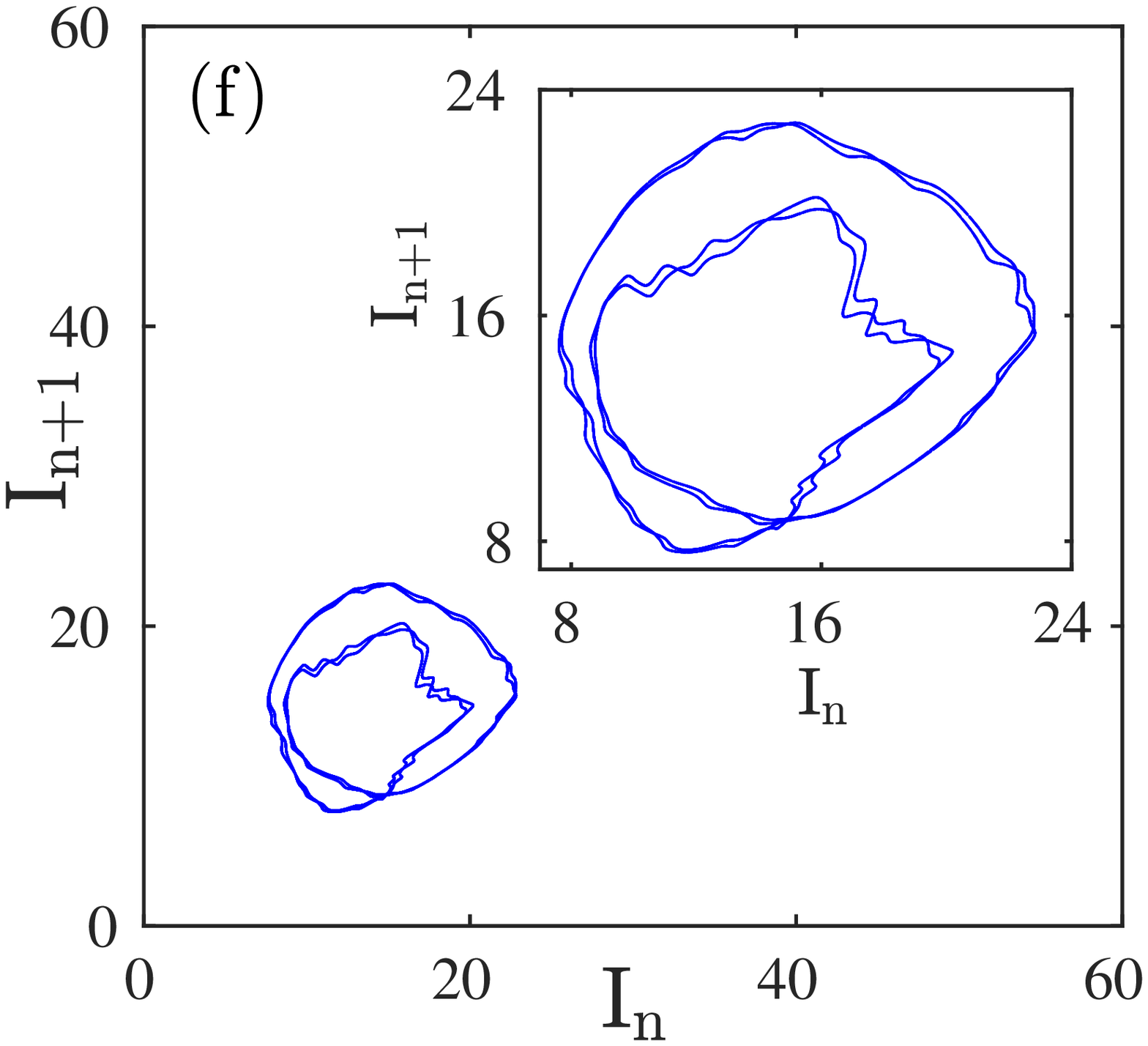} \\
	\includegraphics[width=0.55\columnwidth]{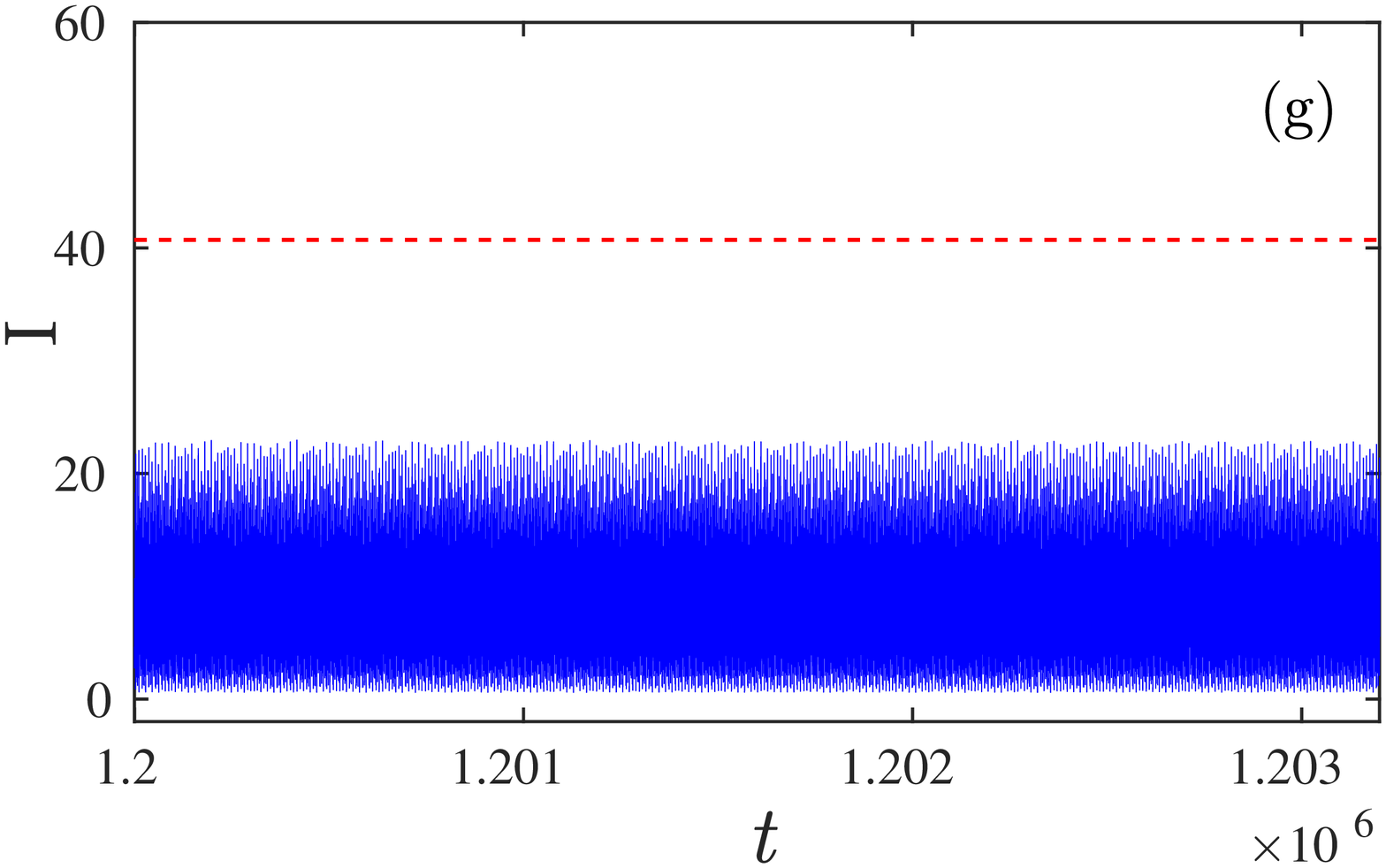}~\includegraphics[width=0.4\columnwidth]{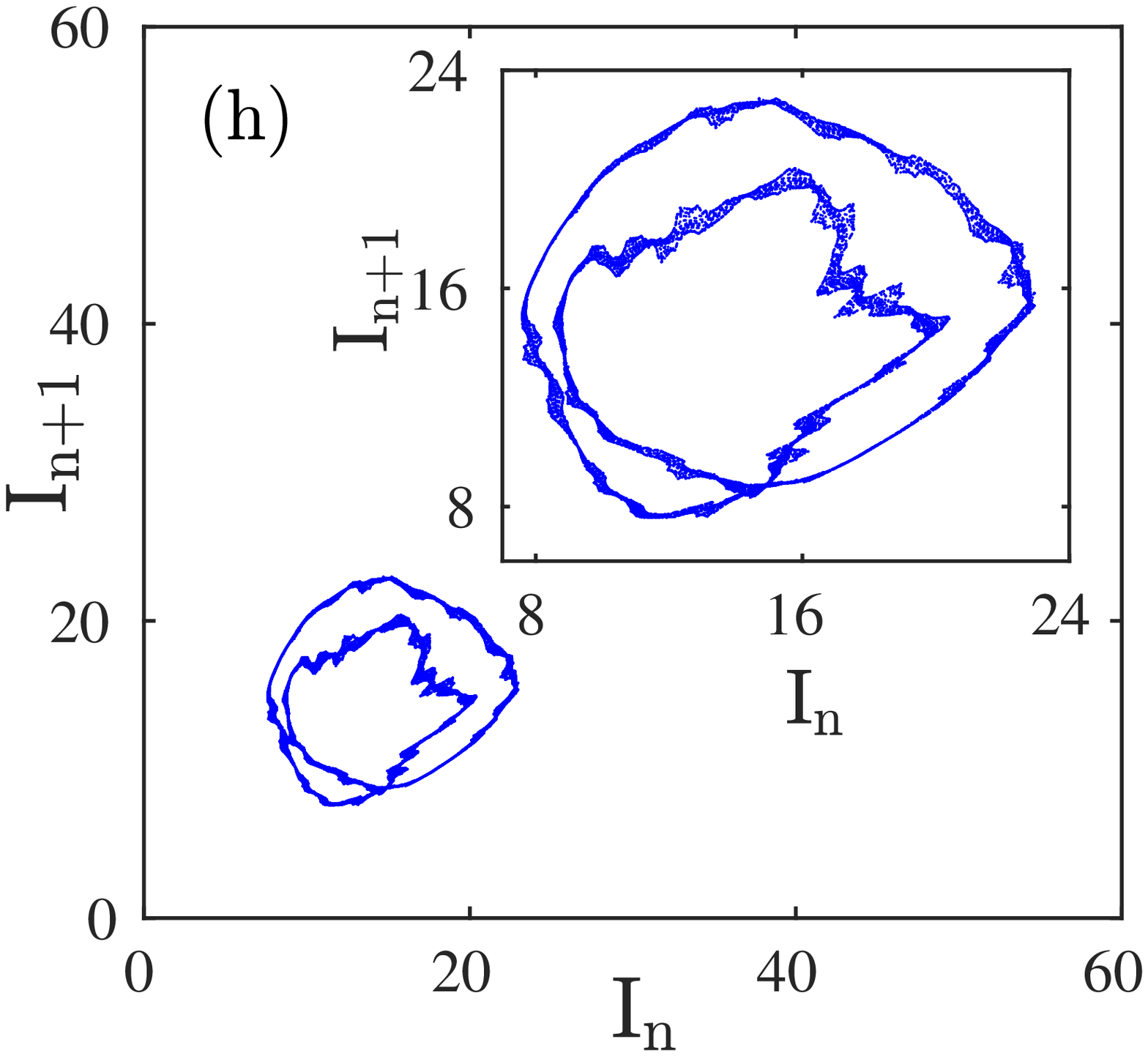} \\
	\includegraphics[width=0.55\columnwidth]{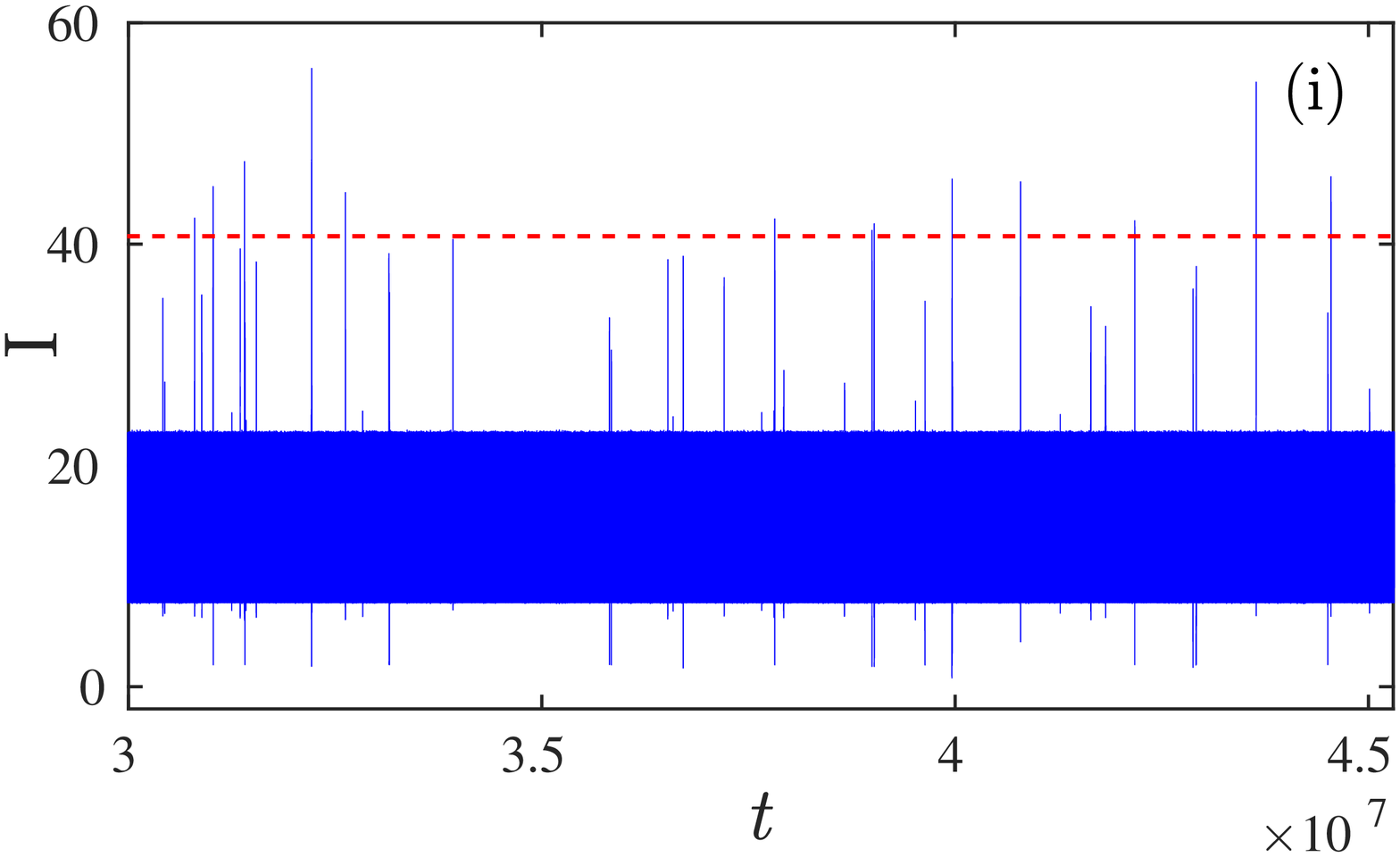}~\includegraphics[width=0.4\columnwidth]{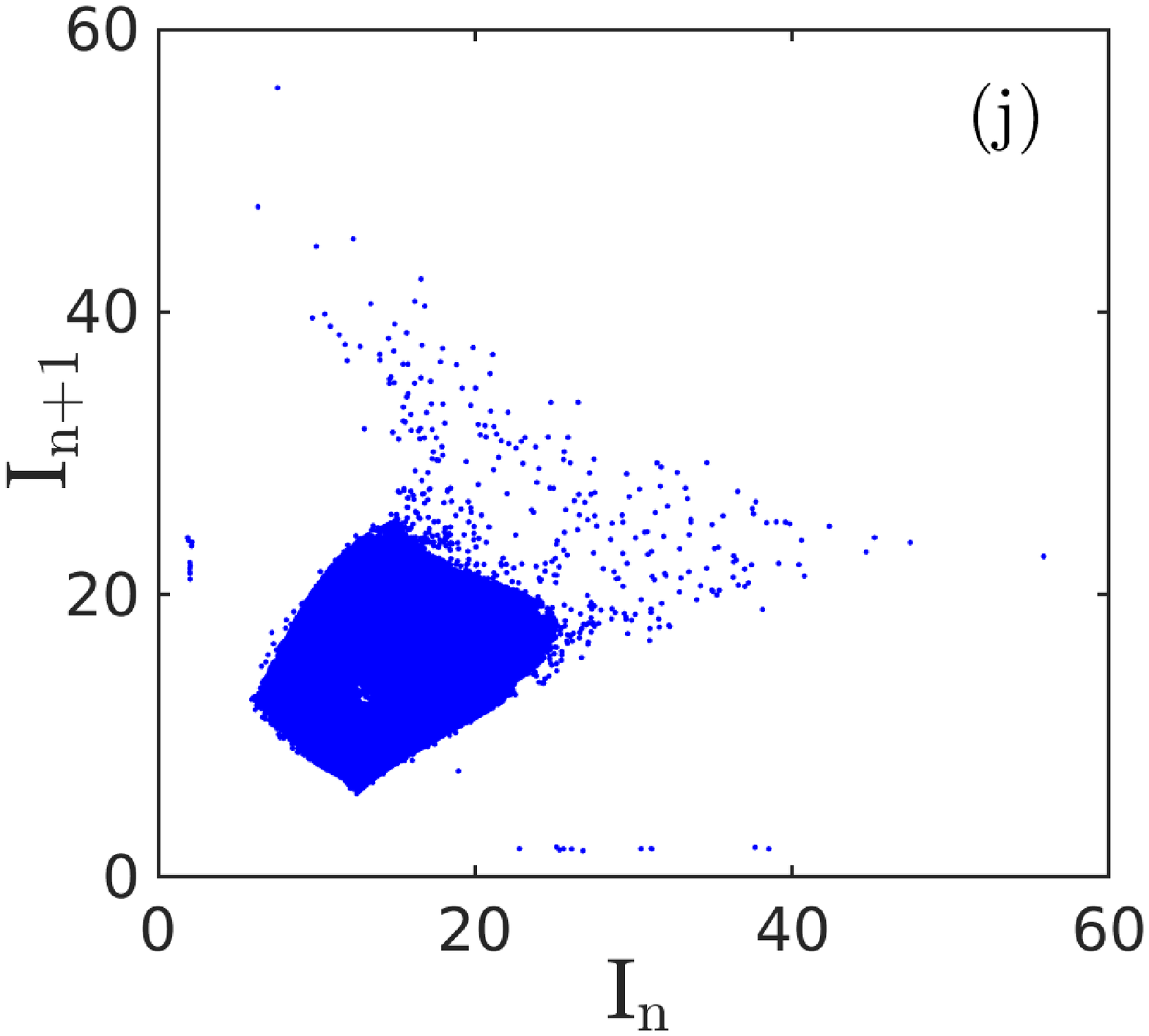} \\
	\caption {Breakdown of quasiperiodic motion via torus-doubling:  Time evaluation of laser intensity (left panels) and its corresponding return maps (right panels). (a, b) One-torus, (c, d) two-torus, (e, f) four-torus, (g, h) chaos, and (i, j) rare large-intensity events for $r$ = 36.74, 35.152, 35.149, 35.1405, and 35.1404, respectively. The horizontal dashed (red) lines in the all time series (panels in the left column) signify $\mathrm{H_s}$=$\langle I_n\rangle$ + $6\sigma_{I}$ . }
	\label{QPB:time_ret}
\end{figure} 
 \section{Large-intensity Pulses: Breakdown of quasiperiodic motion}
A breakdown of QP motion to chaos via a cascade of torus-doubling is a well known phenomenon in nonlinear dynamical systems. This route was demonstrated earlier \cite{Redondo1} in the Zeeman laser model, but the authors ignored the origin of LIE beyond the nominal chaotic mode. However, rare large-intensity pulses emerge in a  range of pumping rate $r$ and are  larger in amplitude or size than a threshold height. We demonstrate here how LIE originates from nominal chaos with a tuning of $r$ in the QPB region and remains indistinguishable in the previous report \cite{Redondo1} since LIE maintains the characteristic feature of nominal chaos with a positive Lyapunov exponent. For this observation, we vary $r$ 
along the horizontal dashed line drawn in Fig.~\ref{Zee:2ph}(b) when $\alpha$ = 7.0 and $\sigma$ = 6.0 are fixed. 
A corresponding bifurcation diagram is drawn in Fig.~\ref{QPB:bif_lya}(a) against $r$ in a range $r \in (35.142, 36.942)$ with local maxima $I_{max}$ of laser intensity $I = E_x^2 + E_y^2$ when we see a transition from periodic to QP motion via Neimark-Sacker bifurcation \cite{Kuznetsov} at a critical $r \approx$ 36.75. The bifurcation diagram is continued for a lower range of $r=(35.133, 35.142)$ in Fig.~\ref{QPB:bif_lya}(b). The transition to chaos is noticed at $r \approx 35.1405$ when the largest Lyapunov exponent $\lambda_1$ shows a transition from zero to a positive value in Fig.~\ref{QPB:bif_lya}(c), yet $I_{max}$ remains bounded in Fig.~\ref{QPB:bif_lya}(b), which suddenly blows up for a little detuning of $r \approx 35.1404$ (marked by a vertical dashed magenta line) when occasional large intensity peaks start appearing. This indicates the origin of LIE, but they are apparently intermittent as seen in  $I_{max}=I_n$ plot in Fig.~\ref{QPB:bif_lya}(b) (rare blue dots) at each $r$ value in the bifurcation diagram. $I_{max}=I_n$ of LIE is larger than a threshold height $\mathrm{H_s}$=$\langle I_n\rangle$ + $6\sigma_{I}$ line (horizontal red line), where $\langle . \rangle$ and $\sigma_I$ denote long time average of $I_n$ and standard deviation, respectively. This characteristic feature of the large event dynamics continues for lower values of $r$, however, $H_s$ is not a constant as shown in Fig.~\ref{QPB:bif_lya}(b), but fluctuates (see inset). 
\par Note that $\lambda_1$ as plotted in Fig.~\ref{QPB:bif_lya}(c) is estimated using a  perturbation method \cite{Balcerzak}, where an integration time of  $2.0\times 10^{7}$ is taken after removing a transient time of $1.0\times 10^{6}$ with a step size $0.01$. At a critical value $r = r_{c} \approx 35.1404$, the system exhibits rare and recurrent LIE ($I_{max}$), which are seen as sudden sparsely  populated dots (blue dots) in Fig.~\ref{QPB:bif_lya}(b). For a specific choice of initial conditions, the Zeeman laser  shows 11 counts of LIE (for the above mentioned time interval) at $r \approx 35.1404$. For decreasing $r$ values, the count (red dots) in Fig.~\ref{QPB:bif_lya}(c) of LIE gradually increases and finally saturates  at $\sim$1100 LIE. 
\par In the range of $r \in (35, 37)$, we find a cascade of torus-doubling, origin of nominal chaos and the transition to LIE as occasional large-intensity chaotic events. Figure~\ref{QPB:time_ret} presents a series of temporal dynamics of laser intensity $I$ (left column) and their return maps in $I_{n+1}$ versus $I_n$ plots (right column) for different $r$. From a visual check of the temporal dynamics of $I$ in Fig.~\ref{QPB:time_ret}(a),  the nature of the dynamics is not clear, however, a closed cycle in the return map  in  Fig.~\ref{QPB:time_ret}(b) confirms the origin of QP motion for $r=36.74$ as seen in Fig.~\ref{QPB:bif_lya}(a). The laser system undergoes a cascade of torus-doubling as shown  in Figs.~\ref{QPB:time_ret}(c) and (d) 
and Figs.~\ref{QPB:time_ret}(e) and \ref{QPB:time_ret}(f) 
when period-2 and period-4 cycles emerge in the return maps for $r=35.152$ and $35.149$, respectively.  Finally, QP motion transits to chaos for a pumping rate $r \approx 35.1405$ as shown in Fig.~\ref{QPB:time_ret}(g)  and confirmed by a  indistinct cycle boundary in the return map in  Fig.~\ref{QPB:time_ret}(h) (inset shows filled-in cycles with a messy boundary). However, laser intensity peaks $I_{max}$ remain restricted to low amplitude, which we define here as nominal chaos. The $H_s$ mark (horizontal dashed red lines) lies far above the instantaneous $I$ value.  
 \par  The temporal evolution of $I$ in Fig.~\ref{QPB:time_ret}(i) confirms a dense boundary of low amplitude events, but accompanied by intermittent very large spiking events and many of them cross the $H_s$ mark (horizontal dashed line) and some of them are even almost three times larger the limit of nominal chaos for $r \approx 35.1404$. The return map shows a dense region (dense blue) in Fig.~\ref{QPB:time_ret}(j) like a comet-head with a tail of rare points (blue dots) scattered at a distance that appears as a dusty cloud. The scattered points denote rare large-intensity pulses called LIE, which are distinctly different from small amplitude nominal chaos and especially, different by their statistical properties.  Noteworthy that 
$E_x^2$ and $E_y^2$ also exhibit LIE when plotted separately, the details of which are presented in the Appendix. 
\begin{figure}
\includegraphics[width=0.5\columnwidth]{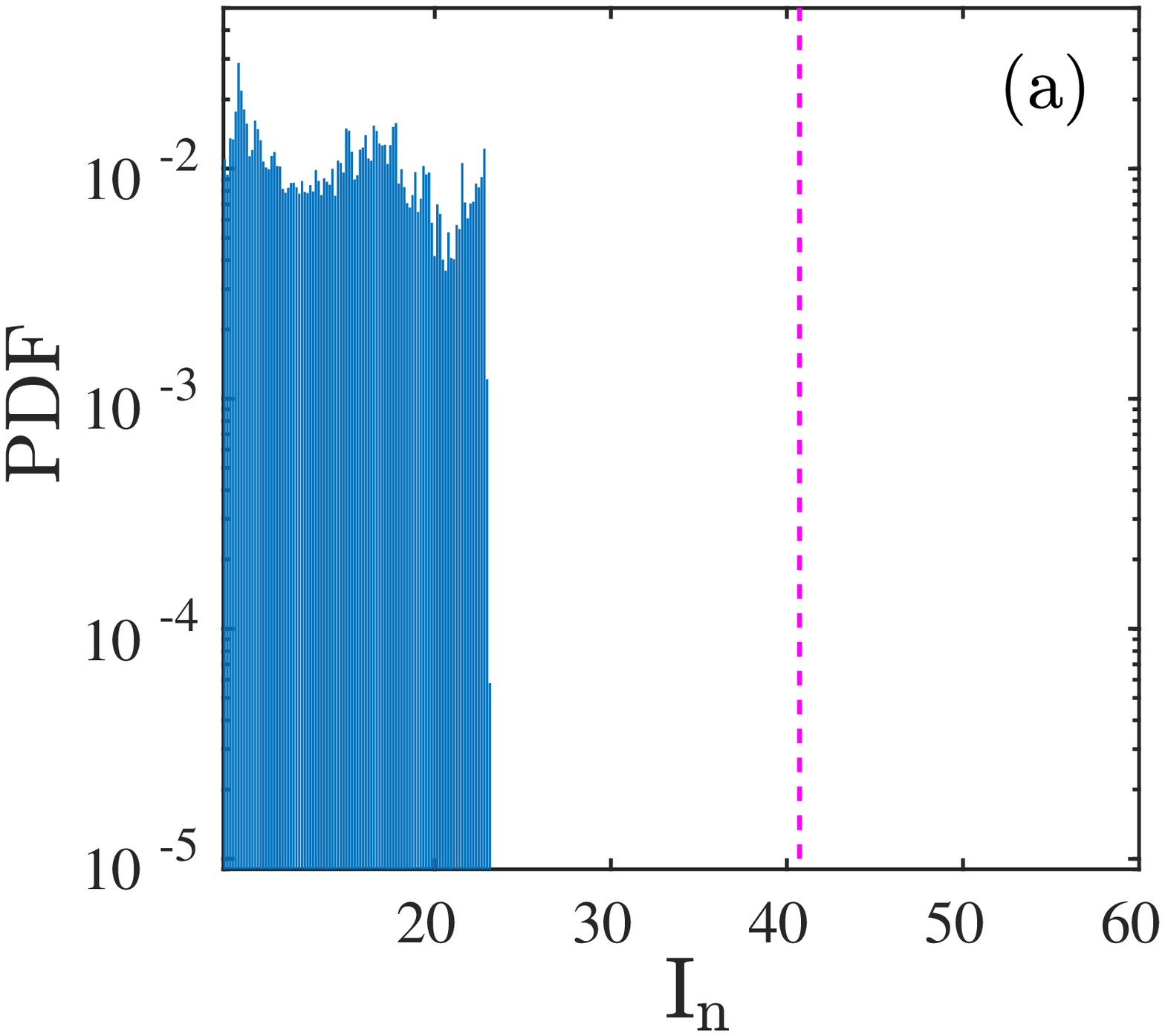}~\includegraphics[width=0.5\columnwidth]{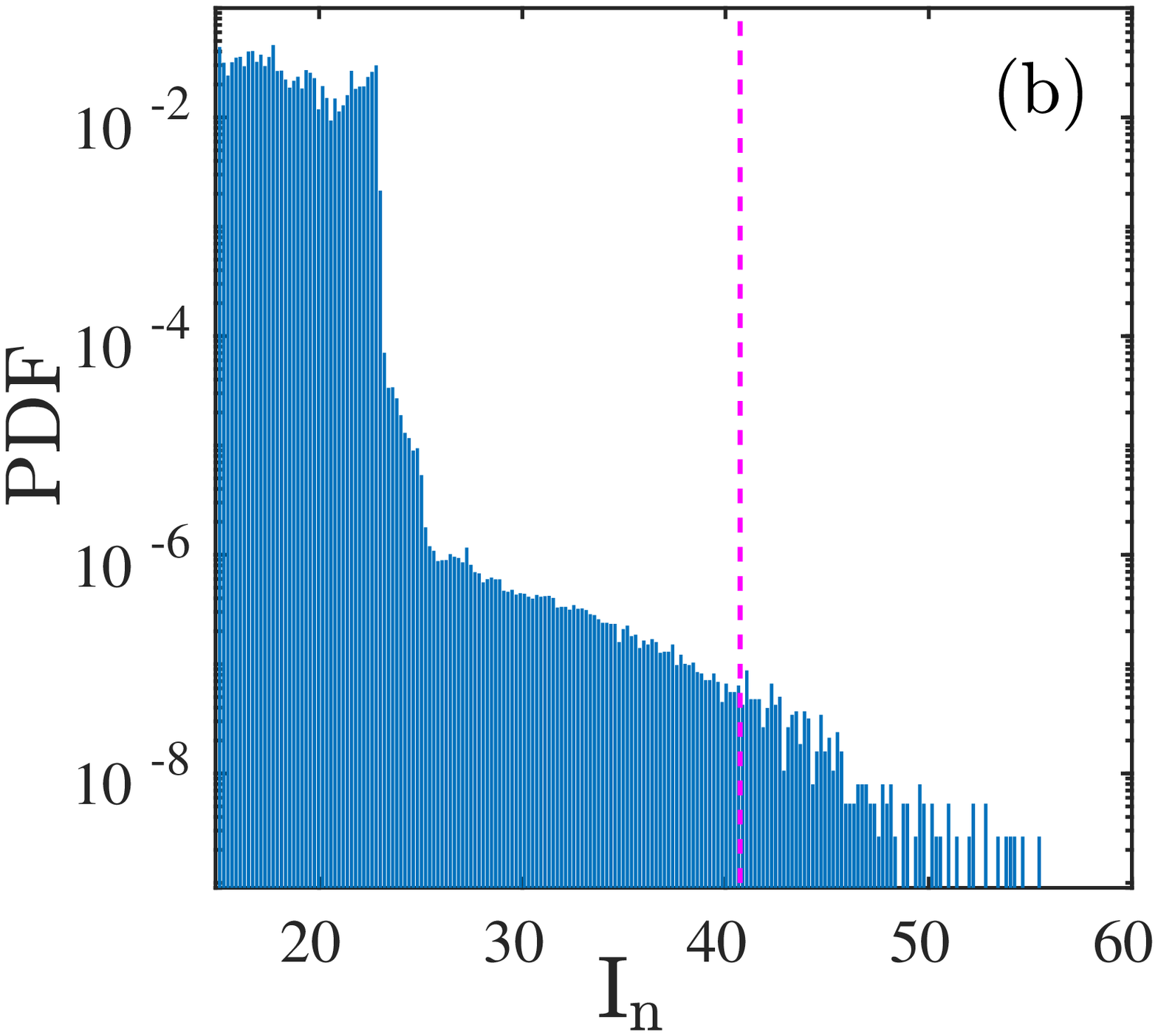}
\caption{Probability distribution function of events. (a) Nominal chaos for $r$ = 35.1405, and (b) LIE for $r$ = 35.1404. Large-intensity events lie in a tail beyond the $H_s$=$\langle I_n\rangle$ + $6\sigma_{I}$ line.}
\label{QPB:pdf}
\end{figure} 
\par The probability distribution function (PDF) of all the peaks $I_{max}=I_n$ against peak size of laser intensity $I_n$ is shown in Fig.~\ref{QPB:pdf}. The distribution in Fig.~\ref{QPB:pdf}(a) for nominal chaos is bounded within a low range of $I_n$ values below the $H_s$ mark (vertical magenta lines) as expected, while it shows non-Gaussian probability distribution of events decaying with the size of events in Fig.~\ref{QPB:pdf}(b) that confirms  low probability of occurrence of LIE beyond the $H_s$ mark (vertical magenta line) for $r=35.1404$. For plotting this PDF, we have taken the $t$-span length as $5.0\times 10^{9}$ after discarding sufficiently long transients, and confirmed that the shape of the distribution does not change with respect to the $t$-span length. 
\section{Large-Intensity pulses: Quasiperiodic intermittency}
\begin{figure}
	\includegraphics[width=1.0
	\columnwidth]{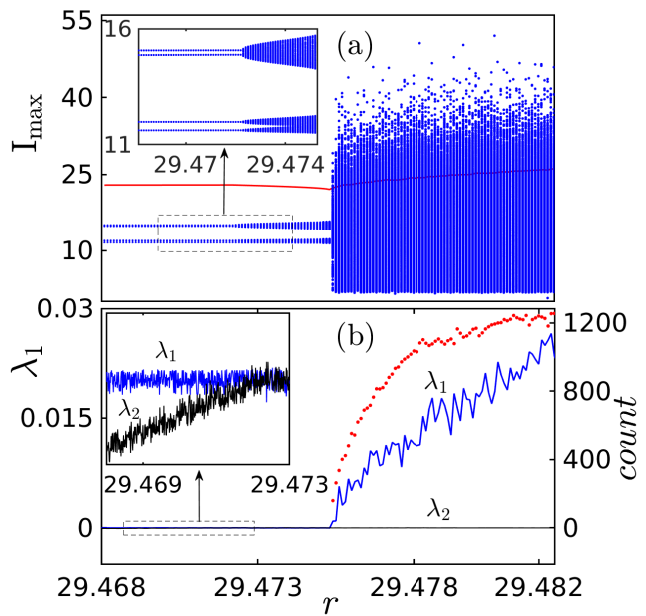} \\
	\caption{(a) Bifurcation diagram of laser intensity against pumping rate $r$. Inset shows a transition from period-4 to QP state. Other parameters are $\sigma$ = 6.0, $\alpha$ = 3.995. (b) Plot of Lyapunov exponents $\lambda_{1,2}$. Inset shows a magnified version of a section marked by a dashed rectangle. Count (red dots) of LIE is plotted against $r$. LIE start appearing at a critical $r=29.4754$.}
	\label{QPI:bif_lya}
\end{figure} 
\begin{figure}
	\includegraphics[width=0.55\columnwidth]{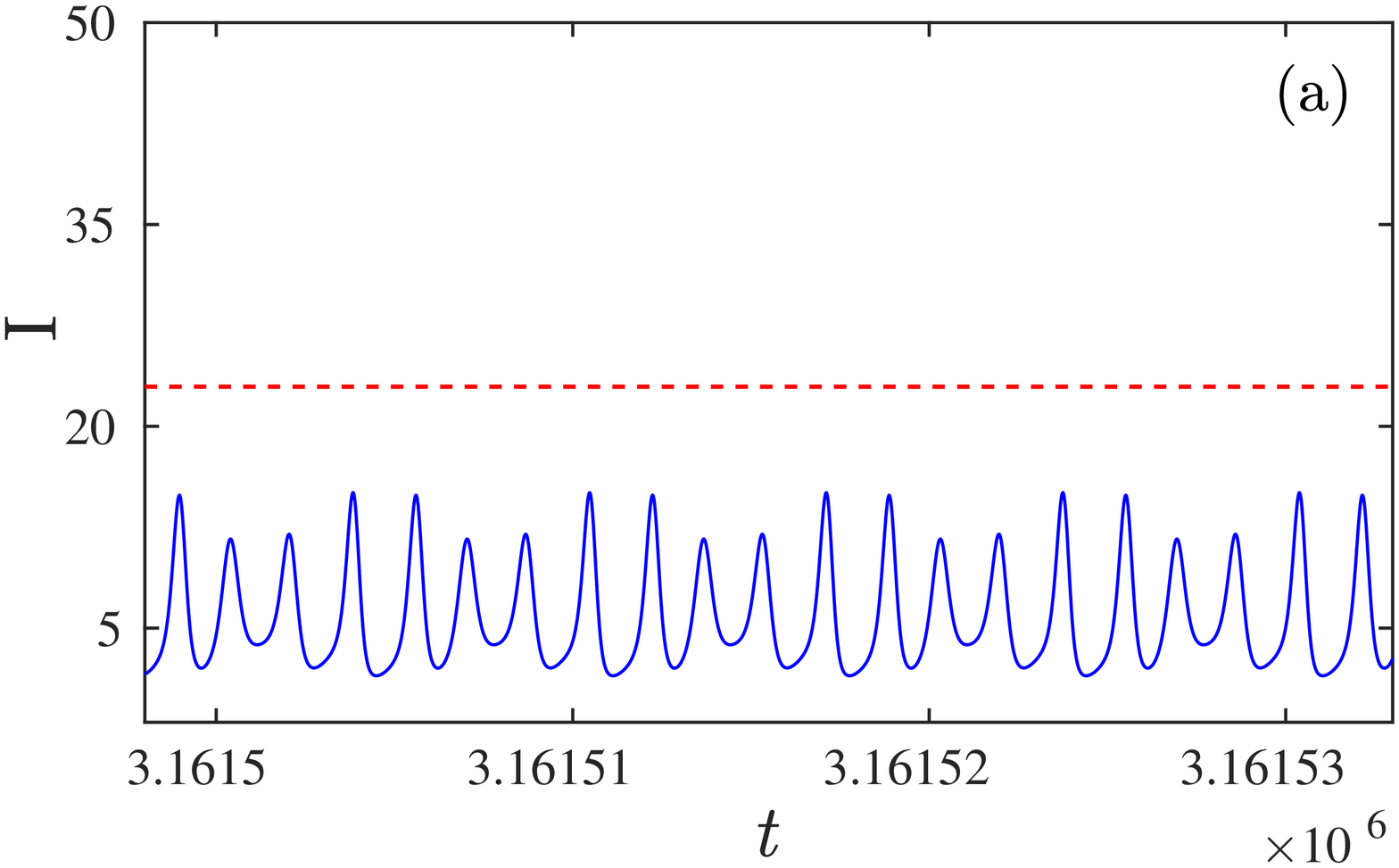}~\includegraphics[width=0.4\columnwidth]{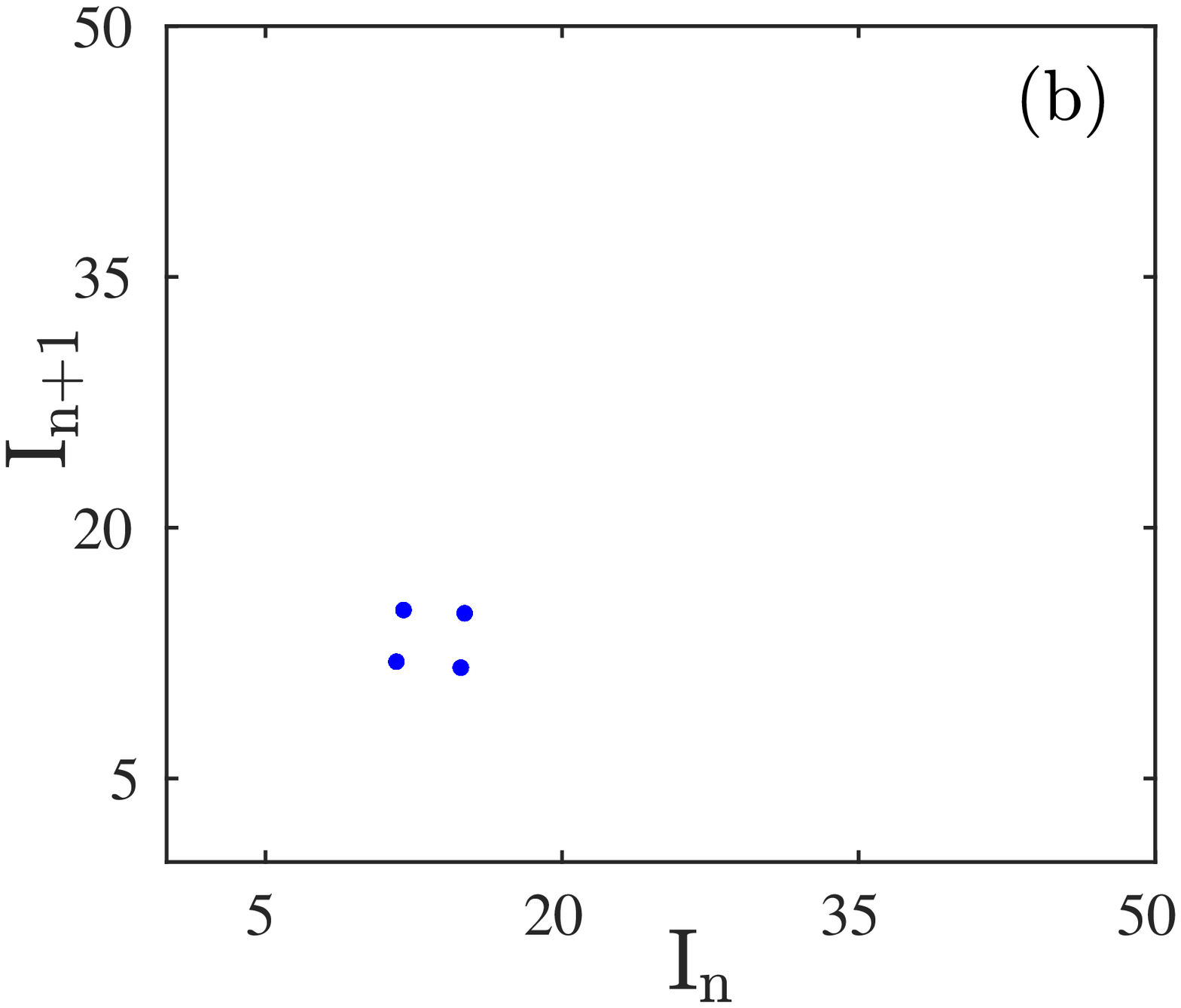}\\
	\includegraphics[width=0.55\columnwidth]{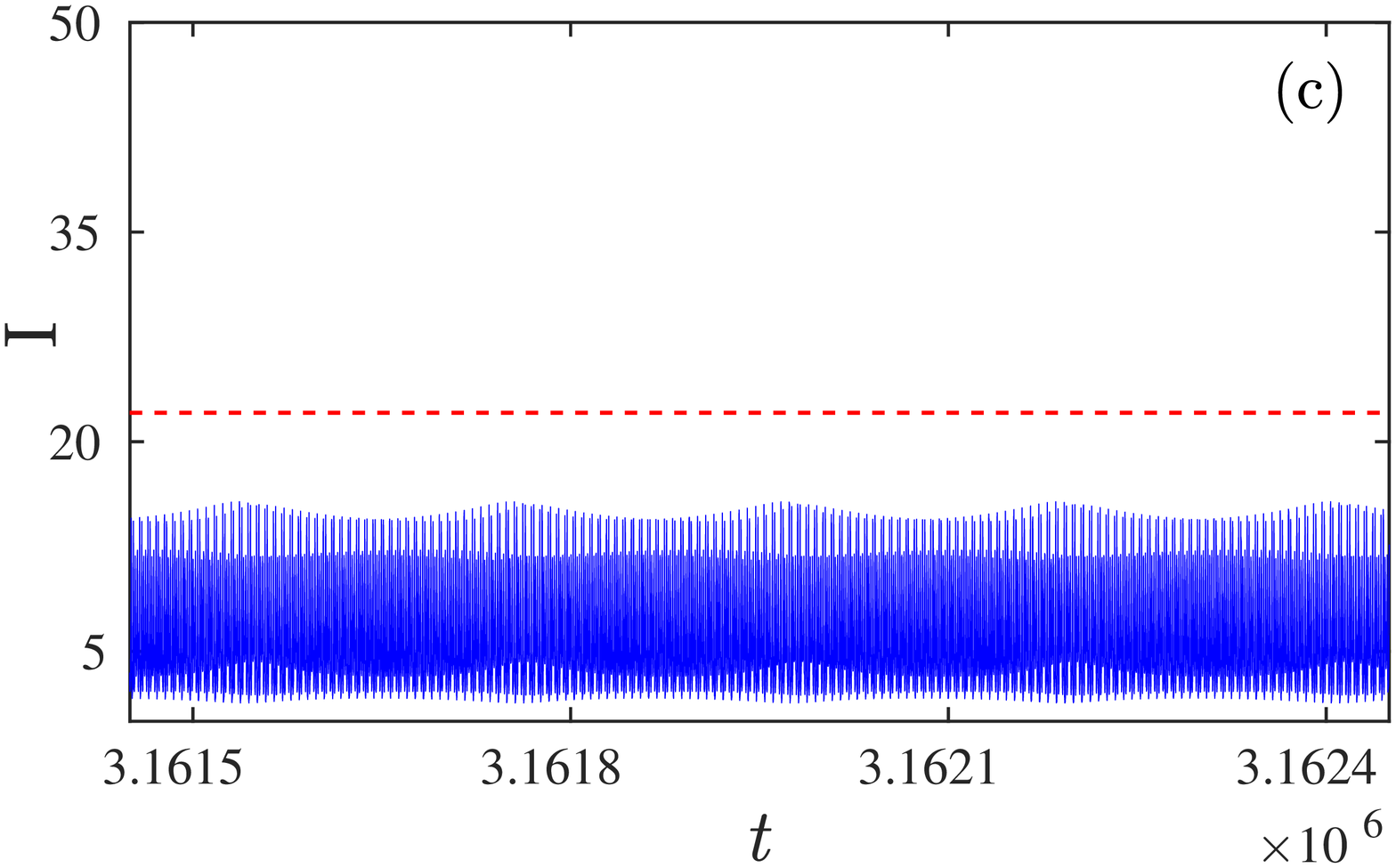}~\includegraphics[width=0.4\columnwidth]{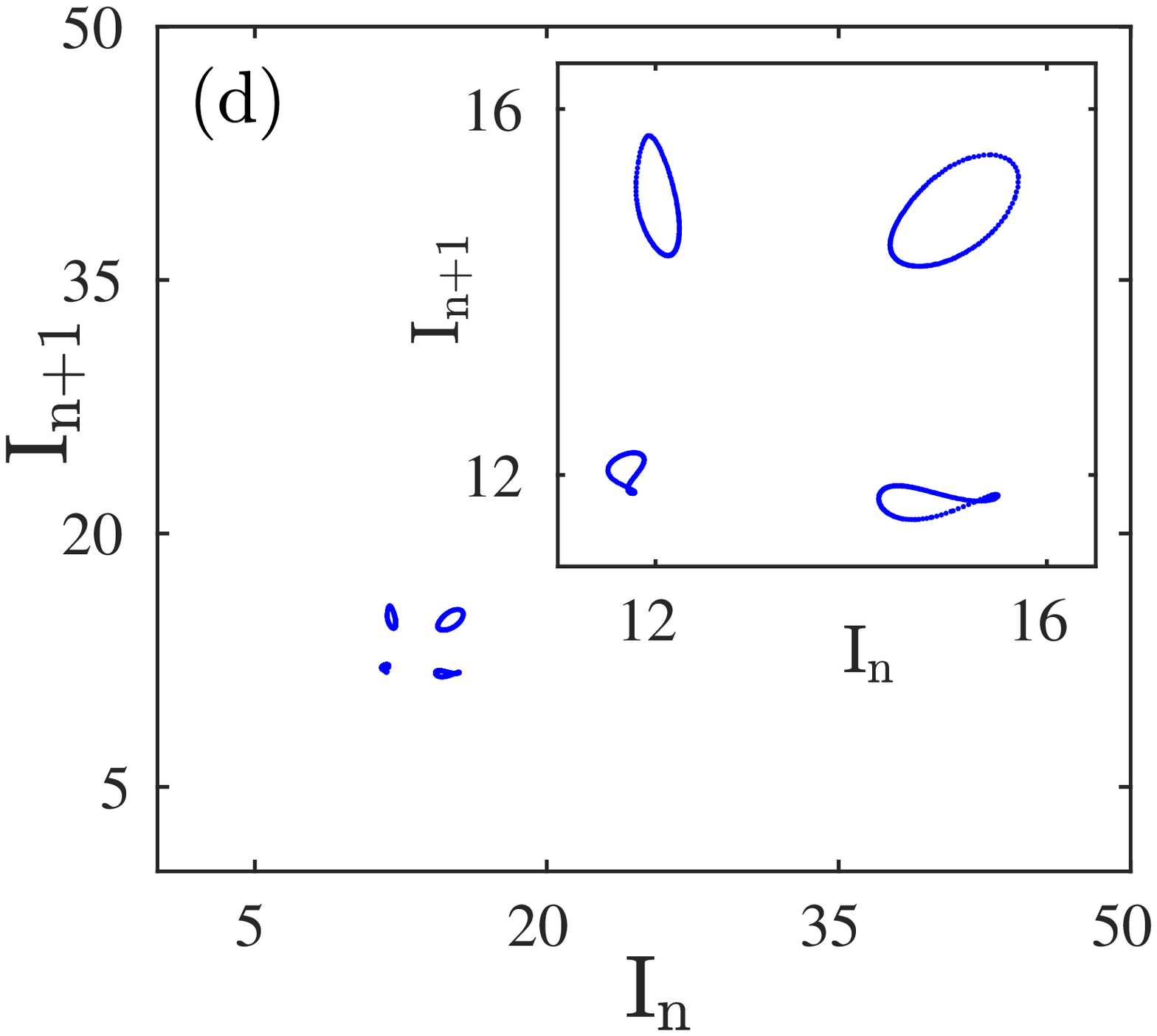}\\
	\includegraphics[width=0.55\columnwidth]{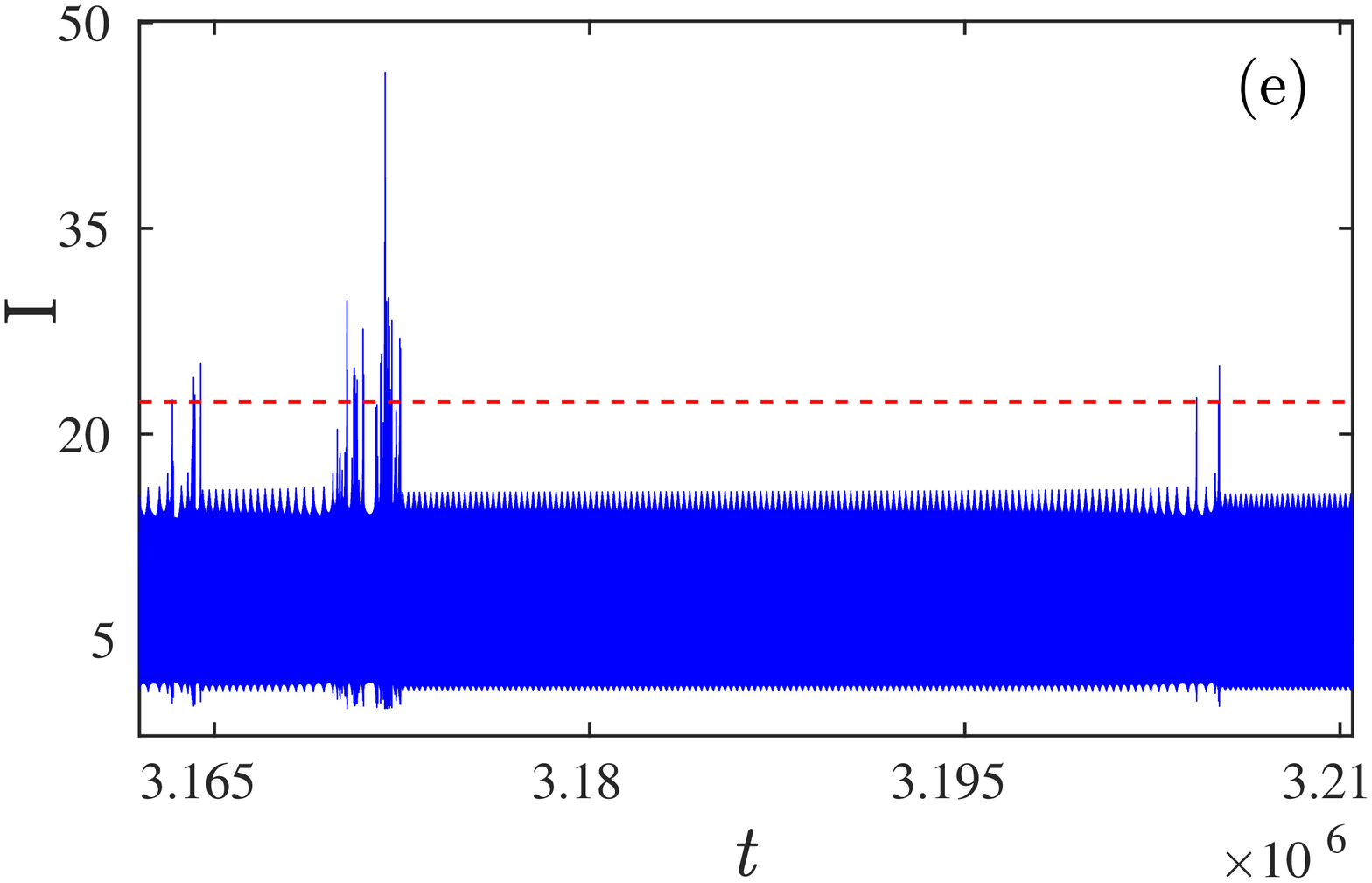}~\includegraphics[width=0.4\columnwidth]{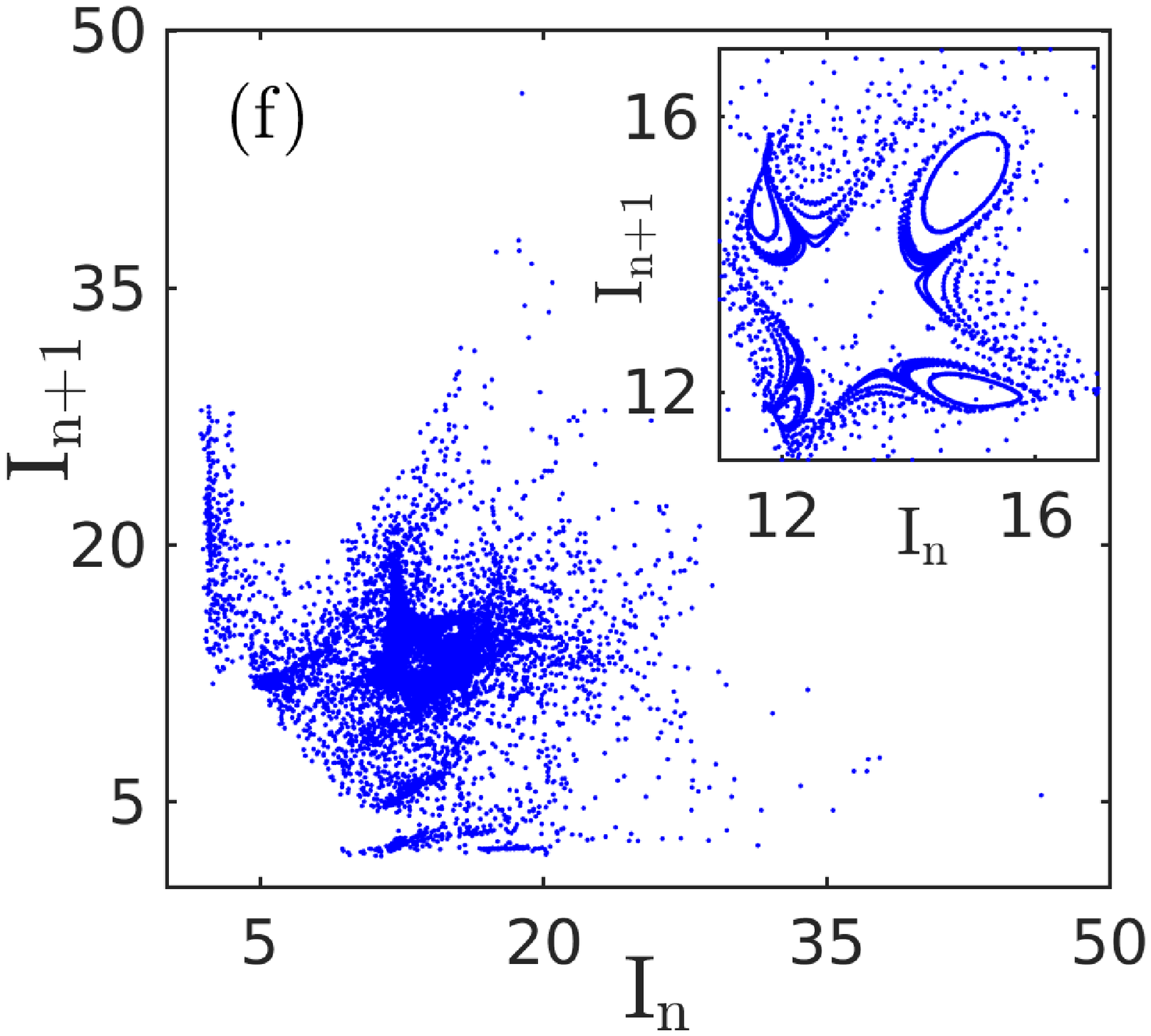} 
	\caption{Temporal evolution of laser intensity (left panels) and return maps (right panels). Period-4 state (a, b), QP motion (c, d), and QP intermittency (e, f) for $r$ = 29.4721, 29.475, and 29.4754, respectively. $H_s$ is marked by horizontal lines in the temporal evolution (red lines).}
	\label{QPI:time_ret}
\end{figure}
Quasiperiodic intermittency is  unusual in dynamical systems, in general, when QP motion is interrupted intermittently by chaotic bursts as found in  Zeeman laser. This unique source of instability leads to occasional LIE as shown here in our numerical experiments. The origin of LIE by QP intermittency was ignored by the authors of the earlier study \cite{Redondo1}. 
For a demonstration of the origin of large events, once again we refer to the phase diagram in Fig.~\ref{Zee:2ph}(c). The dynamics shows a parameter range of periodic state (P, yellow), a region of quasiperiodic state (QP, red), a narrow region of small nominal chaos (C, Blue) and a sea of LIE state (gray). For a better understanding of the transition from one to the other states, we draw a single parameter bifurcation diagram in Fig.~\ref{QPI:bif_lya} against $r$ that follows  the horizontal dashed line in Fig.~\ref{Zee:2ph}(c). Figure \ref{QPI:bif_lya}(a) shows a transition from P (period-4) to QP motion (see inset) then to the LIE state.  A plot  of the largest Lyapunov exponent  $\lambda_1$ (blue line) and the second Large lyapunov exponent $\lambda_2$ (black line) in Fig.~\ref{QPI:bif_lya}(b) confirms the transition from P to QP motion at a critical $r$ $\approx$ 29.4722 when $\lambda_2$ joins $\lambda_1$ at zero (see inset). The laser dynamics finally transits to the LIE state at another critical $r=29.4754$ value when a sudden large increase in amplitude occurs  as shown in Fig.~\ref{QPI:bif_lya}(a). No nominal chaotic state appears here, QP motion directly transits to QP intermittency in a discontinuous manner against $r$. 
The large events start appearing  infrequently as revealed by a  sparse distribution of points (blue dots) above the $H_s$ line (red line). The $H_s$ line is drawn to make a visual impression how large is the size of the LIE. In addition, we plot a count (red dots) of LIE against the pumping rate $r$ in Fig.~\ref{QPI:bif_lya}(b) that increases with $r$ along with increasing $\lambda_1$. LIE counts (red dots) start appearing from the transition point at $r$ = 29.4754. The count of LIE increases monotonically with $r$ and finally saturates at $\approx$ 1200.
\par The transition from period-4 to QP motion and QP intermittency is more clear in Fig.~\ref{QPI:time_ret} where a series of snapshots of temporal dynamics (left panels) and return maps $I_{n+1}$ versus $I_n$ of local maxima $I_{max}=I_n$ (right panels) are displayed for different $r$. The temporal dynamics of laser intensity in Fig.~\ref{QPI:time_ret}(a) and the $I_{n+1}$ versus $I_n$ return map in Fig.~\ref{QPI:time_ret}(b) for $r = 29.4721$, confirm period-4 oscillation. The return map shows four distinct points as a clear indicator of period-4 oscillation. The period-4 oscillation becomes QP motion  for a larger $r = 29.475$ as shown in Fig.~\ref{QPI:time_ret}(c). This is confirmed by its return map in Fig.~\ref{QPI:time_ret}(d), where four distinct cycles  evolve from four distinct points in Fig.~\ref{QPI:time_ret}(b). For a larger $r \approx 29.4754$, QP motion transits to QP intermittency, which is apparent from the time evolution in Fig.~\ref{QPI:time_ret}(e). QP motion is interrupted by occasional chaotic bursts: a typical signature of intermittency except that the laminar phase is now quasiperiodic. Rare large-intensity spikes are seen during the chaotic bursting that are much larger than the $H_s$ line (red line). Rare large-intensity pulses  are reflected in the return map in Fig.~\ref{QPI:time_ret}(f) as scattered  points (blue dots) far from a densely populated central region (dense blue region). The dense region is centered around the four cycles of the quasiperiodic motion as shown in the inset of Fig.~\ref{QPI:time_ret}(f).
\begin{figure}
	\includegraphics[width=0.9\columnwidth]{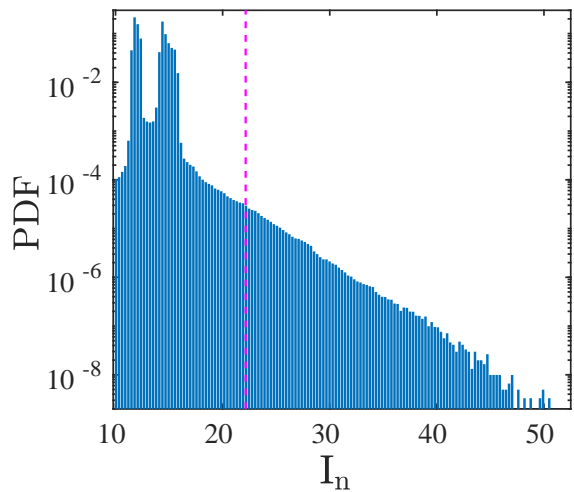}
	\caption{Probability distribution function of intensity peaks during QP intermittency in Zeeman laser. The distribution of laser intensity peaks is non-Gaussian and extended beyond the $H_s$ threshold (vertical dashed line) with a decaying probability with larger $I_n$ for $r$ = 29.4754.} 
	\label{QPI:pdf}
\end{figure} 
\par  PDF for all the events ($I_{max}=I_n$) during QP intermittency is shown in Fig.~\ref{QPI:pdf}. The distribution is non-Gaussian and slowly decays (a heavy tail) with increasing height of large events beyond the $H_s$ line (vertical magenta line). 

\section{conclusion}
Origin of chaos via period-doubling followed by crisis-induced-intermittency and PM intermittency are common sources of instabilities that may originate intermittent large-intensity pulses in lasers. Two other nonlinear processes that lead to chaos, were reported earlier \cite{Redondo1} in the Zeeman laser model, namely, breakdown of QP motion via torus-doubling and QP intermittency. While the breakdown of QP motion via torus-doubling and origin of chaos is well known, in the literature, the QP intermittency is unusual in dynamical systems, in general. We revisited the dynamics of the Zeeman laser model, especially focused on the parameter regions of the sources of instability leading to chaos. Our study has been extended beyond chaos with rigors of numerical simulations, using phase diagrams in two-parameter plane and one parameter bifurcation diagrams and, focuses on the smaller range of  parameters where transition to chaos occurs. Two reasonably significant parameter regions are found where the dynamics is distinctly different from nominal chaos. The nominal chaos is defined here as bounded in amplitude below a well defined threshold height. When the pumping rate of the Zeeman laser is extended beyond the nominal chaos, intermittent large-intensity pulses emerge denoted here as LIE. This information was missing in the previous report \cite{Redondo1}, may be because LIE have same characteristic feature of chaos as having positive Lyapunov exponent. Yet LIE are distinct by their height larger than a threshold and their occasional departure from nominal chaos as seen in a long observation. 
\par In one region of parameter space, LIE appears beyond nominal chaos that emerges via torus-doubling of QP motion.  An interior-crisis is possibly involved during the transition from chaos to the origin of LIE, which needs further rigors of study to confirm.  From the extreme events' perspectives, origin of intermittent large pulses via instability of QP motion is not so common although reported earlier \cite{Mishra_2018} in a coupled neuron model under repulsive synaptic interactions. In another region of parameters, LIE originate via QP intermittency that is, particularly, new and not reported so far, in other dynamical systems, to the best of our knowledge. Another distinct feature of LIE is their  probability distribution in an observable that shows an extended tail beyond the threshold height. In contrast, PDF is bounded below the threshold height for nominal chaos.

\begin{acknowledgments}
T.K., S.L.K., and A.M. have been supported by the National Science Centre, Poland, OPUS Programme Project No. 2018/29/B/ST8/00457. S.L.K acknowledges PLGrid Infrastructure (Poland) for computation facilities.  S.K.D. acknowledges financial support from the Division of Dynamics, Lodz University of Technology, Poland. M.B. is supported by the National Science Centre, Poland, Project No. 2017/27/B/ST8/01619.\\
\end{acknowledgments}  
\section{appendix}
 
\begin{figure}
	\includegraphics[width=0.75\columnwidth]{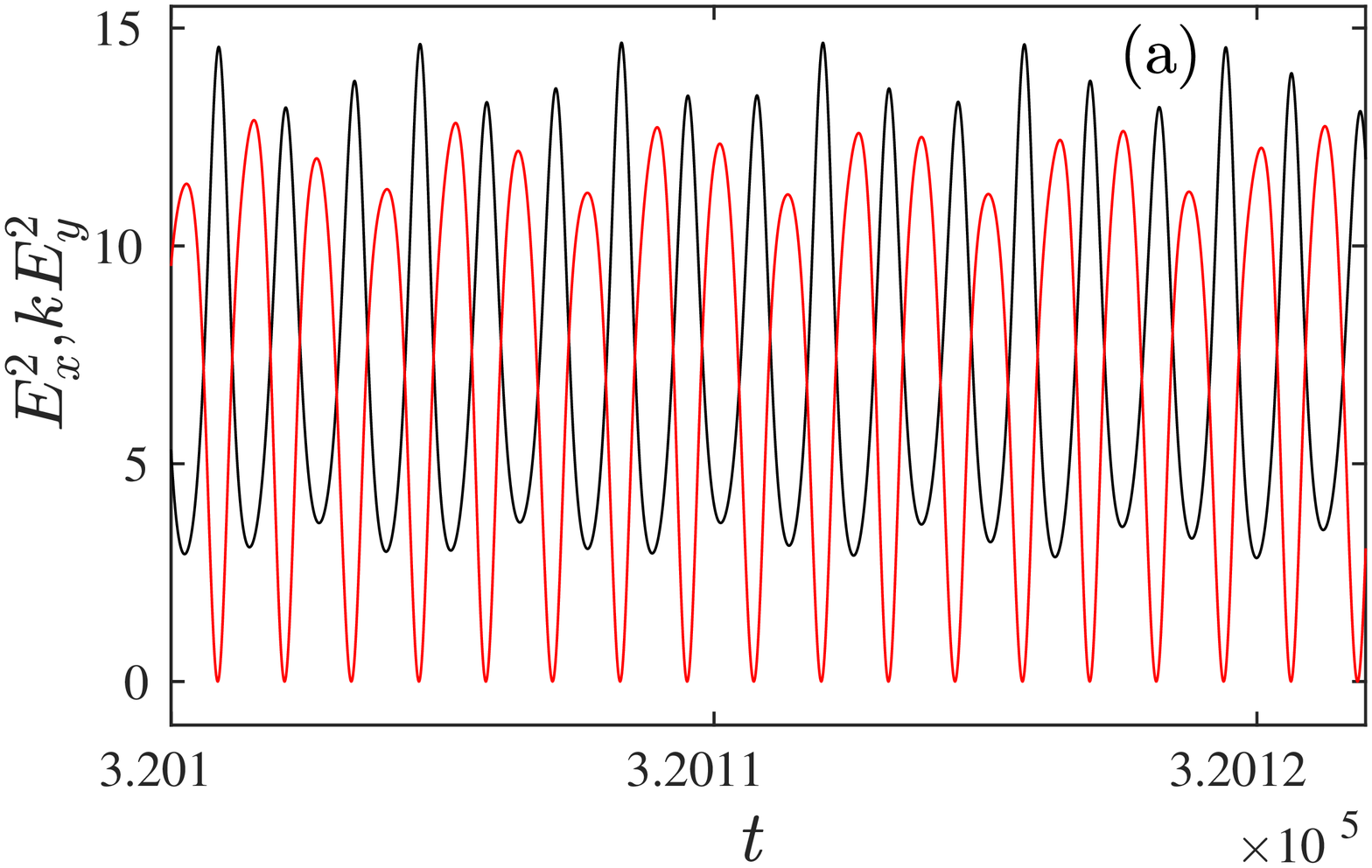}\\
	\includegraphics[width=0.75\columnwidth]{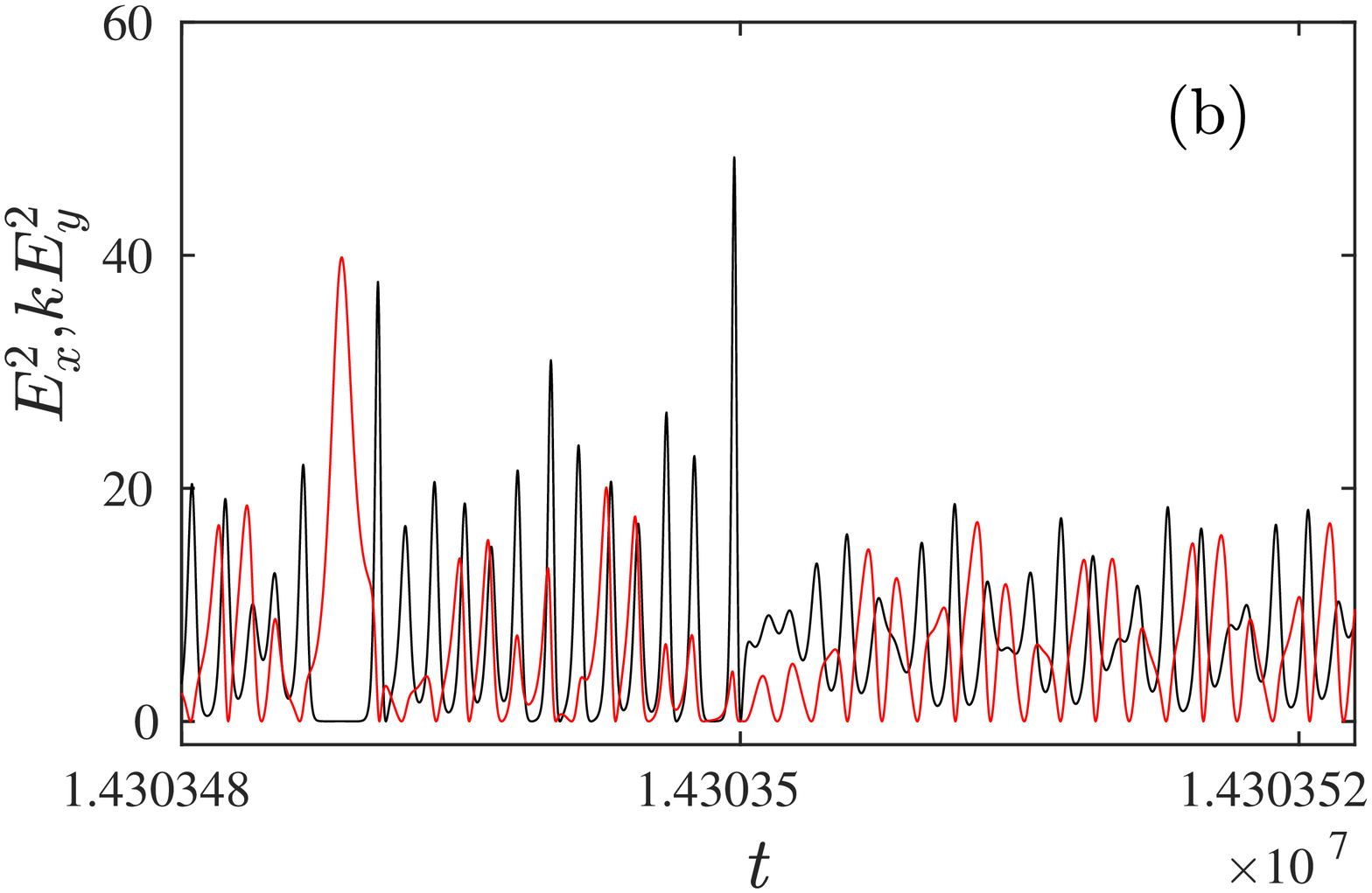} 
	\caption{QPB route to LIE in Zeeman laser model. Temporal dynamics of $E_x^2$ (black line) and $kE_y^2$ (red line), (a) in antiphase QP motion for $r$ = 36.74 and no LIE, (b) LIE are prominent in  $E_x^2$ plot (black line) for $r$ = 35.1404 when the antiphase  correlation  with $E_y^2$ (red line) is lost.  $k= 20$ (arbitrarily chosen) for an enhanced visual comparison against $E_x^2$.}
	\label{QPB_PER:dyn}
\end{figure}

We confirm here that all the dynamical and statistical characteristics of $I$ are present in $E_x^2$ and $E_y^2$ as illustrated here with their separate observation. The temporal dynamics of  $E_x^2$ and $E_y^2$ for fixed  $\alpha$ = 7.0, $\sigma$ = 6.0 are presented here separately for two different $r$ values. For $r$ = 36.74, the laser exhibits QP mode of oscillation  in Fig.~\ref{QPB_PER:dyn}(a) where the temporal dynamics of $E_x^2$ (black line) and $E_y^2$ (red line) manifest antiphase correlation. Since $E_y^2$ is much smaller in size, we have scaled up $E_y^2$ by $k$ $=20$ (arbitrarily chosen) to make it comparable in size with $E_x^2$ for an enhanced visualization. 
For a pumping rate $r$ = 35.1404 when the Zeeman laser exhibits LIE, 
the antiphase correlation between two signals $E_x^2$ and $E_y^2$ breaks down as shown in Fig.~\ref{QPB_PER:dyn}(b). The signature of intermittent large-intensity pulses is present in both the signals, however, it is dominantly present in $E_x^2$, in particular, with larger size by approximately $k=20$-fold compared to $E_y^2$. To identify the LIE, we have to define two new threshold heights as denoted by $h_{s_x, s_y}$=$\langle m_{x,y}\rangle +6\sigma_{x,y}$, where super-subscripts ($x, y$) represent ($E_x^2$, $E_y^2$) signals, and $m_x$ is the local maxima of $E_x^2$, $m_y$ is the local maxima of $E_y^2$, $\sigma_{x,y}$ are their corresponding standard deviations. 
\begin{figure}
	\includegraphics[width=0.5\columnwidth]{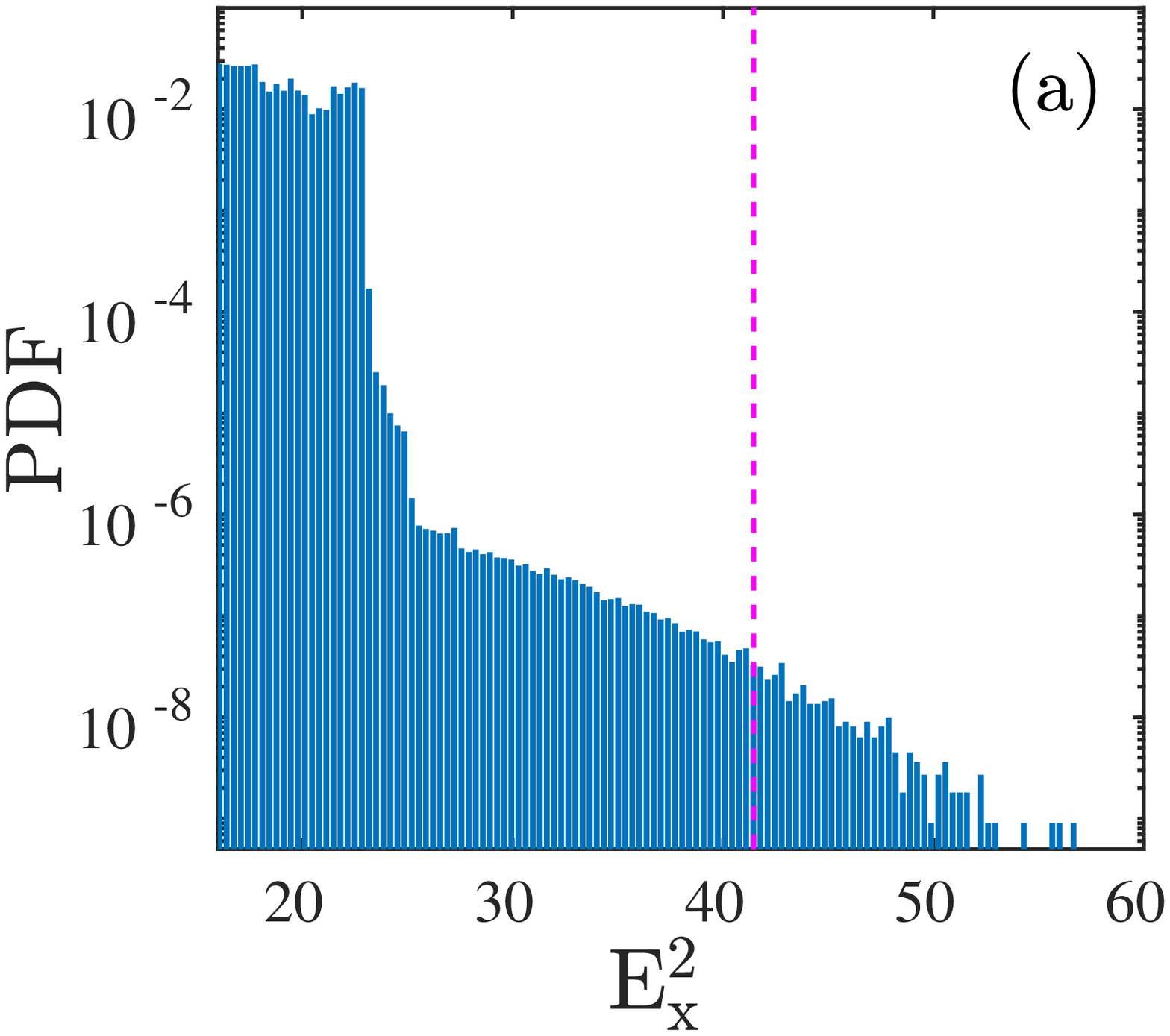}~\includegraphics[width=0.5\columnwidth]{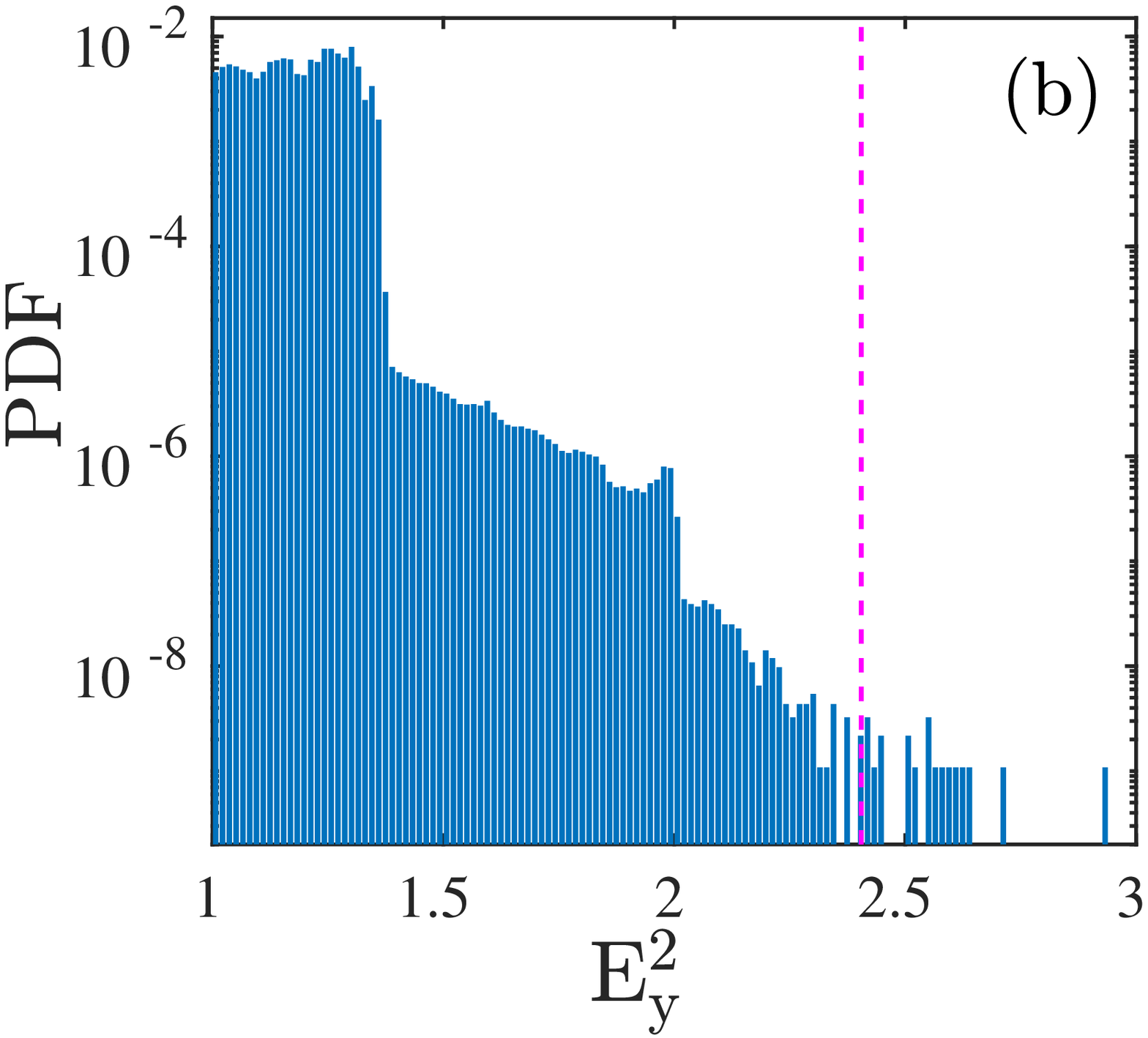} 
	\caption{Quasiperiodic breakdown route to LIE. Probability distribution function of $E_x^2$ (a), and $E_y^2$ (b) for $r$ = 35.1404 that depict extended decaying distributions beyond their respective threshold height $h_{s_x, s_y}$  (dashed vertical lines).}
	\label{Ex_Ey:pdf}
\end{figure} 
PDF of $E_x^2$ and $E_y^2$ are displayed in Figs.~\ref{Ex_Ey:pdf}(a) and \ref{Ex_Ey:pdf}(b), respectively, and both of them show non-Gaussian and extended distribution with a decreasing probability of occurrence of LIE beyond their respective threshold height (dashed vertical line). PDFs are similar to Fig.~\ref{QPB:pdf}(b).
 \begin{figure}
 	\includegraphics[width=0.75\columnwidth]{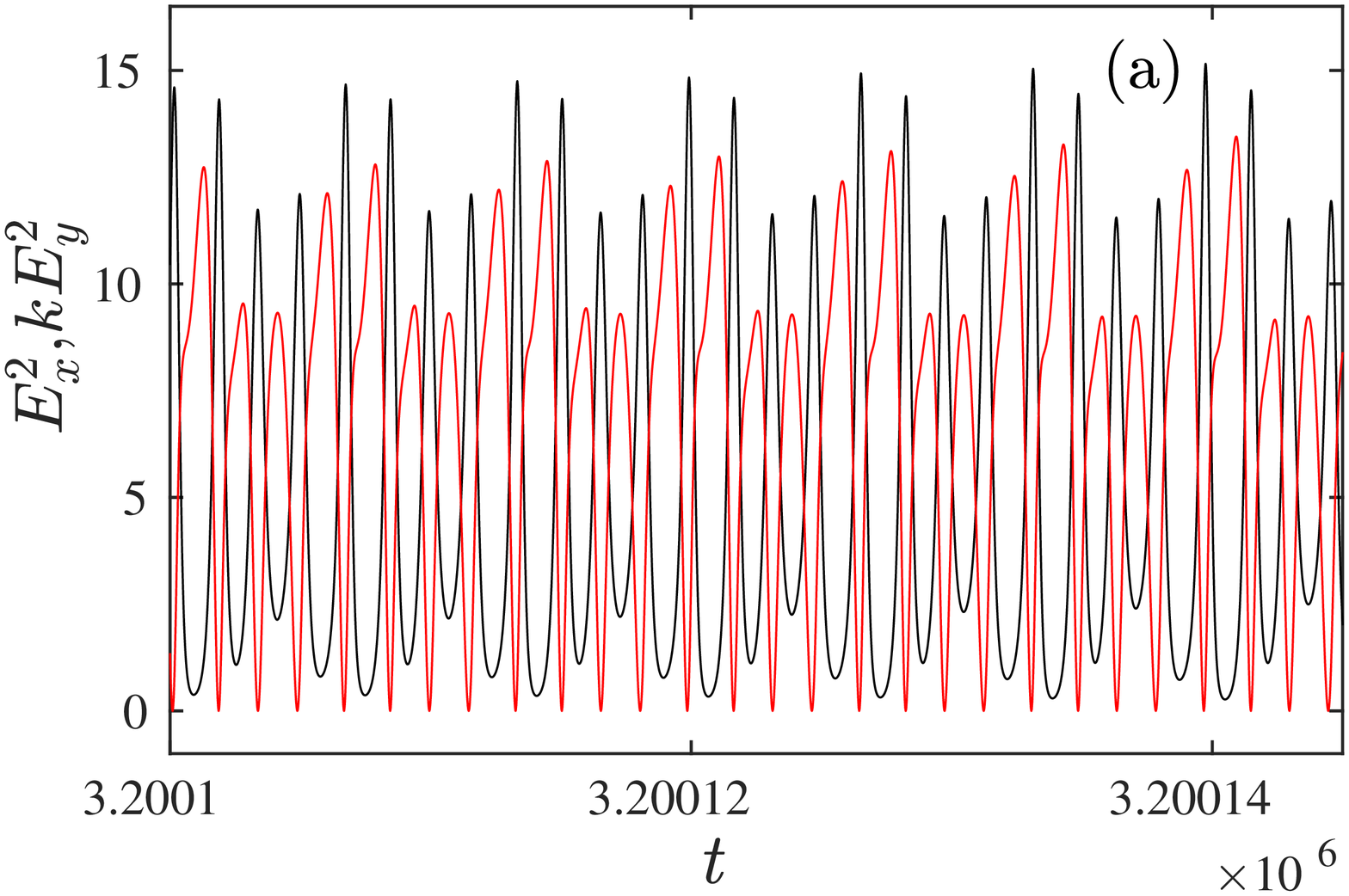}\\
 	\includegraphics[width=0.75\columnwidth]{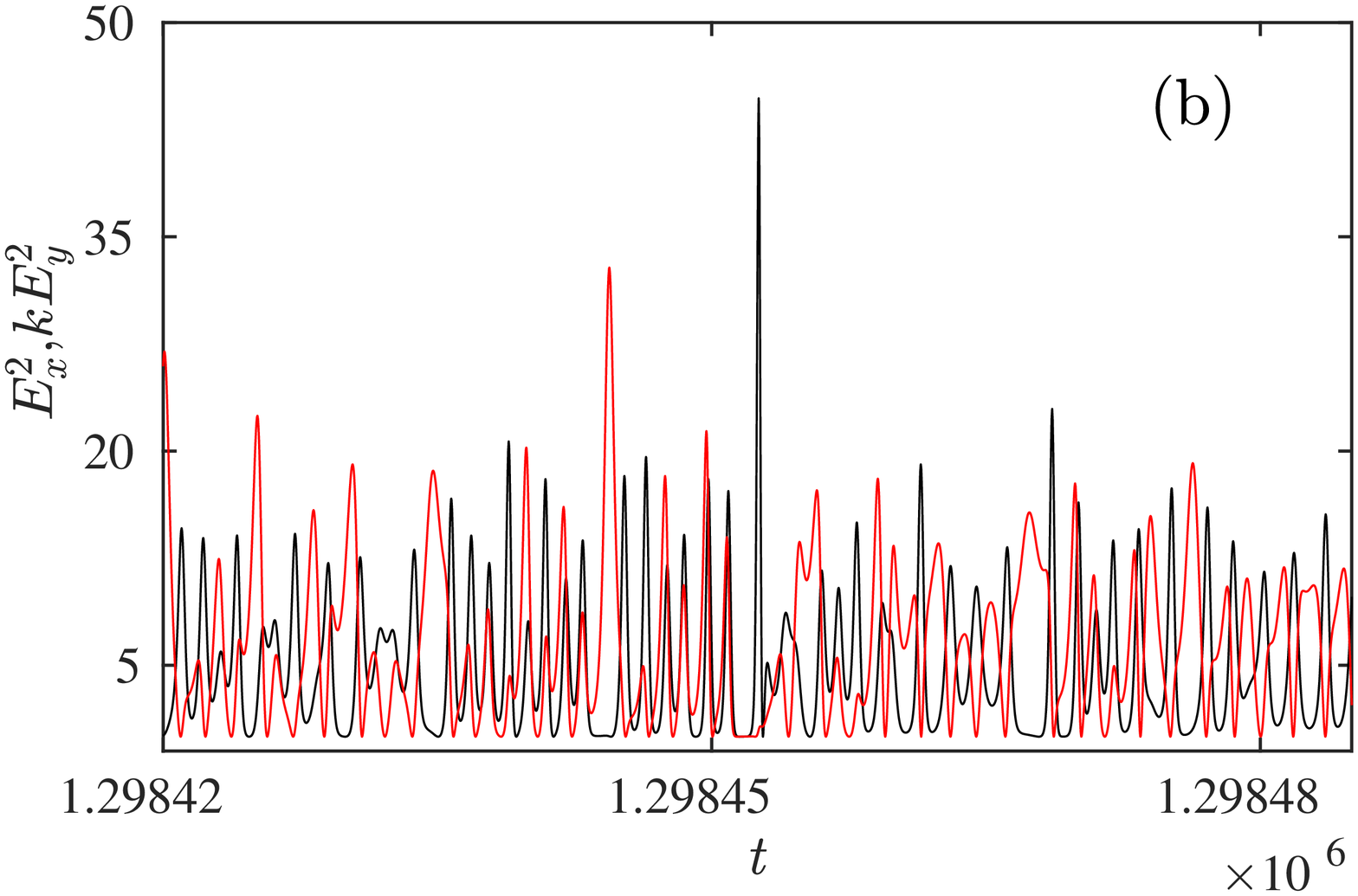} 
 	\caption{ QPI route to LIE in the Zeeman laser model.  Time evolution of $E_x^2$ (black line), and $kE_y^2$ (red line) in  quasiperiodic antiphase state (a) for $r$= 29.475, and LIE (b) for $r$ = 29.4754, where $k=8$ (arbitrarily chosen) for a better visualization. The antiphase correlation is lost during LIE.}
 	\label{QPIPER:dyn}
 \end{figure}
\par We show the temporal evolution of $E_x^2$ and $E_y^2$ for the QPI case in Fig.~\ref{QPIPER:dyn}(a) for $r$ = 29.475 (QP state), and  Fig.~\ref{QPIPER:dyn}(b) for  $r$ = 29.4754 (LIE). Other parameters are  $\alpha$ = 3.995 and $\sigma$ = 6.0. It is clear that  $E_x^2$ and $E_y^2$ and manifest antiphase correlation during QP  motion once again and when LIE originate, the antiphase relation is lost.  For this case, we arbitrarily scaled up $E_y^2$ by  eight times here in both Figs.~\ref{QPIPER:dyn}(a) and \ref{QPIPER:dyn}(b) as done for the QPB case. PDFs of both $E_x^2$ and $E_y^2$ follow the same trend as shown in Fig.~\ref{QPI:pdf} for $I_n$ and hence we decide as redundant for presentation here.   
 

\begin{thebibliography}{100}
 
\bibitem{Arecchi} F. T. Arecchi, and R. G. Harrison, (Eds.) \emph{Instabilities and Chaos in Quantum Optics} (Springer, Berlin, 1987).
\bibitem{Arecchi1} J. R. Tredicce, F. T. Arecchi, G. L. Lippi, and G. P. Puccioni, \textit{J. Opt. Soc. Am.} B {\bf 2}, 173 (1985).
\bibitem{Arecchi2} F. T. Arecchi, Phys. Scr. {\bf T23}, 160 (1988).
\bibitem{Roy} K. S. Thornburg, Jr., M. M\"oller,  R. Roy, T. W. Carr, R.-D. Li, and T. Erneux, Phys. Rev. E {\bf 55}, 3865 (1997).
\bibitem{Junji} J. Ohtsubo,  \emph{Semiconductor lasers: Stability, Instability and Chaos} (Springer, Berlin, 2006).
\bibitem{PM} 
Y.Pomeau, and P. Manneville, Comm. Math. Phys. {\bf 74}, 189 (1980).
\bibitem{Roy1} F. Rogister, and R. Roy, Phys. Rev. Lett. {\bf 98}, 104101 (2007).
\bibitem{Solli} D. R. Solli, C. Ropers, P. Koonath, and B. Jalali, Nature 
{\bf 450}, 1054 (2007).
\bibitem{Bonatto}
C. Bonatto, M. Feyereisen, S. Barland, M. Giudici, C. Masoller, Jose R. R. Leite, and J. R. Tredicce, Phys. Rev. Lett. \textbf{107}, 053901 (2011).
\bibitem{Pisarchik} A. N. Pisarchik, R. Jaimes-Re\'ategui, R. Sevilla-Escoboza, G. Huerta-Cuellar, and M. Taki, Phys. Rev. Lett. {\bf 107}, 274101 (2011).
\bibitem{Montina}
A. Montina, U. Bortolozzo, S. Residori, and F. T. Arecchi, Phys. Rev. Lett. \textbf{103}, 173901 (2009).
\bibitem{Granese}
N. M. Granese, A. Lacapmesure, M. B. Ag\"uero, M. G. Kovalsky, A. A. Hnilo, and J. R. Tredicce, Opt. Lett. \textbf{41}, 3010-3012 (2016).

\bibitem{Rimoldi}
C. Rimoldi,  S. Barland, F. Prati, and G. Tissoni, Phys. Rev. A \textbf{95}, 023841 (2017).
\bibitem{Cristian-CO2} C. Bonatto, and A. Endler, Phys. Rev. E {\bf 96}, 012216 (2017).
\bibitem {Mercier} \'E. Mercier, A. Even, E. Mirisola,  D. Wolfersberger, and M. Sciamanna, Phys. Rev. E, $\bf 91$, 042914 (2015).
\bibitem{Dysthe} K. Dysthe, H. E. Krogstad, and P. Müller, Annu. Rev. Fluid Mech. \textbf{40}, 287 (2008).
\bibitem{Arnob_Phys_Report} S. Nag Choudhury,  A. Ray, S. K. Dana, and D. Ghosh, ``Extreme events in dynamical systems and random walkers: A review" (unpublished).
\bibitem{Reinoso} J. A. Reinoso, J. Zamora-Munt, and C. Masoller, Phys. Rev. E $\bf87$, 062913 (2013).
\bibitem{Zamora-Munt} J. Zamora-Munt, B. Garbin, S. Barland, M. Giudici, Jose R. R. Leite, C. Masoller, and J. R. Tredicce, Phys. Rev. A $\bf87$, 035802 (2013).
\bibitem{Metayer} C. Metayer, A. Serres, E. J. Rosero, W. A. S. Barbosa, F. M. de Aguiar, J. R. Rios Leite, and J. R. Tredicce, Opt. Express {\bf 22}, 19850 (2014).
\bibitem{Grebogi} C. Grebogi, E. Ott, F. Romeiras, J. A. Yorke, Phys. Rev. A {\bf 36}, 5365 (1987); C. Grebogi, E. Ott, and J. A. Yorke, Physica D {\bf 7}, 181 (1983).
\bibitem{Rakshit} A. Roy, S. Rakshit, D. Ghosh, and S. K. Dana, Chaos {\bf 29}, 043131 (2019).
\bibitem{kingston} S. L. Kingston, K. Thamilmaran, P. Pal, U. Feudel, and S. K. Dana, Phys. Rev. E \textbf{ 96}, 052204 (2017).
\bibitem{Coulibaly}	 
S. Coulibaly, M. G. Clerc, F. Selmi, and  S. Barbay, Phys. Rev. A \textbf{95}, 023816 (2017).	

\bibitem{Clerc1}
M. G. Clerc, G. Gonz\'alez-Cort\'es, and M. Wilson, Opt. Lett. {\bf 41}, 2711  (2016).

\bibitem{Clerc} F. Selmi, S. Coulibaly, Z. Loghmari, I. Sagnes, G. Beaudoin, M. G. Clerc, and S. Barbay, Phys. Rev. Lett. {\bf  116}, 013901 (2016).
\bibitem{Lucerini} V. Lucarini, D. Faranda, J. M. M. de Freitas, M. Holland, T. Kuna, M. Nicol, M. Todd, and S. Vaienti, \emph{Extremes and Recurrence in Dynamical Systems} (Wiley, New York, 2016).
\bibitem{deOliveira} G. F. de Oliveira, O. Di Lorenzo, T. P. de Silans, M. Chevrollier, M. Ori\'a, and Hugo L. D. de Souza Cavalcante, Phys. Rev. E \textbf{93}, 062209 (2016).


\bibitem{Suresh_2018} K. Suresh, and A.N. Pisarchik, Phys. Rev. E \textbf{98}, 032203 (2018).

\bibitem{Karnatak_2014} R. Karnatak, G. Ansmann, U. Feudel, and K. Lehnertz, Phys. Rev. E \textbf{90}, 022917 (2014).
\bibitem{Ott} H. L. D. de Souza Cavalcante,  Marcos Ori\'a, D. Sornette, E. Ott, and D. J. Gauthier, Phys. Rev. Lett. {\bf 111}, 198701 (2013).
\bibitem{Mishra_2018}	A. Mishra, S. Saha, M. Vigneshwaran, P. Pal,  T. Kapitaniak, and S.K.  Dana, Phys. Rev. E \textbf{97}, 062311 (2018).
\bibitem{Ray_2019} A. Ray, A. Mishra, D. Ghosh, T. Kapitaniak,  S. K. Dana,  and C. Hens, Phys. Rev. E \textbf{101}, 032209 (2020).
\bibitem{Sayantan} S. Nag Chowdhury, S. Majhi, M. Ozer, D. Ghosh, and M. Perc, New J. Phys {\bf 21}, 073048 (2019).
\bibitem{Majhi} S. Majhi, S. Nag Chowdhury and D. Ghosh, Euro. Phys. Letts.  {\bf 132}, 20001 (2020).
\bibitem{Sayantan1} S. Nag Chowdhury, S. Majhi, D. Ghosh, IEEE Trans. Netw. Sci. Engg. {\bf 7}, 3159 (2020).
\bibitem{Chang} W. Chang, J. M. Soto-Crespo, P. Vouzas, N. Akhmediev,  Opt. Lett. \textbf{40}, 2949 (2015).	
\bibitem{Mohamad} M. A. Mohamad, T. P. Sapsis, Ocean Eng. \textbf{120}, 289 (2016).

\bibitem{Arnob} A. Ray, S. Rakshit, G. K. Basak, S. K. Dana, and D. Ghosh, Phys. Rev. E 101, 062210 (2020).
\bibitem{Farazmand} M. Farazmand,  T. P. Sapsis, Sci. Adv. \textbf{3}, 1701533 (2017); M. Farazmand T. Sapsis, 	Appl. Mech. Rev.\textbf{ 71}, 050801 (2019).
\bibitem{kingston2020}
S. L. Kingston, K. Suresh, K. Thamilmaran,  and T. Kapitaniak, Eur. Phys. J. Special Topics \textbf{229}, 1033 (2020).

\bibitem{Mishra2020}A. Mishra, S. L. Kingston, C. Hens, T. Kapitaniak,  U. Feudel, and S. K Dana, Chaos, \textbf{30},  063114 (2020).


\bibitem{Lehnertz} K. Lehnertz, ``Epilepsy: Extreme events in the human brain" in \emph{Extreme Events in Nature and Society} (Springer, Heidelberg, 2005), p. 123.

\bibitem{Nicolis} C. Nicolis, V. Balakrishnan, and G. Nicolis, Phys. Rev. Lett. \textbf{97}, 210602 (2006).
\bibitem{Puccioni} G. P. Puccioni, M. V. Tratnik, and J. E. Sipe, Opt. Lett. {\bf 12}, 242 (1987).
\bibitem{Abraham} N. B. Abraham, M. D. Matlin, and R. S. Gioggia, Phys. Rev A {\bf 53},3514 (1996).
\bibitem{Redondo}  J. Redondo, E. Rold\'an, and G. J. de Valc\'arcel, Phys. Lett. A {\bf 210}, 301 (1996).
\bibitem{Redondo1} J. Redondo, G. J. de Valc\'arcel, and E. Rold\'an, Phys. Rev. E, \textbf{56}, 6589 (1997).
\bibitem{Kuznetsov} Y. A. Kuznetsov, \emph{Elements of Applied Bifurcation Theory} (Springer, New York, 1998).
\bibitem{Balcerzak}
M. Balcerzak,  D. Pikunov, and A. Dabrowski, Nonlin. Dynam. \textbf{94}, 3053 (2018).

\end{thebibliography}
\end{document}